\documentclass[12pt]{article}
\usepackage{amssymb,amsmath,epsfig}

\begin{document}

\title{\bf Dynamics of Scalar Thin-Shell for a Class of Regular Black Holes}
\author{M. Sharif$^1$ \thanks{msharif.math@pu.edu.pk} and
Sehrish Iftikhar $^{1,2}$ \thanks{sehrish3iftikhar@gmail.com}\\
$^1$ Department of Mathematics, University of the Punjab,\\
Quaid-e-Azam Campus, Lahore-54590, Pakistan.
\\$^2$Department of Mathematics, Lahore College\\
for Women University, Lahore-54000, Pakistan.}
\date{}
\maketitle
\begin{abstract}
This paper is devoted to study the dynamical behavior of thin-shell
composed of perfect fluid by considering matter field as a scalar
field. We formulate equation of motion of the shell by using Israel
thin-shell formalism for a class of regular black holes as interior
and exterior regions. The corresponding scalar fields and effective
potentials are investigated numerically for both massless and
massive scalar fields. We conclude that massless scalar shell leads
to collapse, expansion and equilibrium while the massive case leads
to collapse only.
\end{abstract}

\section{Introduction}

Scalar fields play a key role in several astrophysical phenomena and
have many applications in theoretical physics, cluster dynamics and
cosmology. Wheeler (1955) found particle like solutions (geons) from
classical electromagnetic field coupled to general relativity (GR)
which were extended by Brill and Wheeler (1957) as well as by
Frederick (1957). Bergmann and Leipnik (1957) investigated solution
of the Einstein field equations in the presence of scalar field for
Schwarzschild geometry. Christodoulou (1991) examined spherically
symmetric scalar collapse and formation of singularities. Choptuik
(1993) studied spherically symmetric collapse of a massless scalar
field minimally coupled with gravity and discussed its solutions
numerically. Harada et al. (1997) investigated collapsing compact
object in scalar-tensor theory which predicts the existence of
scalar gravitational waves.

The study of dynamics of an astrophysical object with scalar field
has been the subject of keen interest for many people. Kaup (1968)
as well as Ruffini and Bonazzola (1969) were the pioneers to study
boson stars (composed of self-gravitating complex scalar field).
Gleiser (1988) investigated the dynamical instability of boson stars
and concluded that instabilities do not occur for the critical
centra1 density but for centra1 densities considerably higher.
Seidel and Suen (1990) examined dynamical evolution of boson stars
using perturbation and discussed the existence as well as formation
of boson stars. Siebel et al. (2001) studied dynamics of massless
scalar field interacting with a self-gravitating neutron star and
found that the scalar wave forces the neutron star either to
oscillate or to undergo gravitational collapse to form a black hole
(BH) on a dynamical timescale. Bhattacharya et al. (2009) explored
spherically symmetric collapse of a massless scalar field and
discussed possibility for the existence of a class of nonsingular
models.

Chase (1970) examined instability of spherically symmetric charged
fluid shell by using equation of state. Boulware (1973) investigated
time evolution of the charged thin-shell and found that end state of
collapse could be a naked singularity if and only if density is
negative. Barrab$\grave{e}$s and Israel (1991) studied dynamics of
thin-shells traveling at the speed of light. $N\acute{u}\tilde{n}ez$
et al. (1997) investigated stability and dynamical behavior of a
real scalar field for the Schwarzschild BH. Goncalves (2002)
examined dynamical properties of timelike thin-shell by using Israel
formalism and formulated necessary and sufficient condition for the
shell crossing. Oren and Piran (2003) explored evolution of a
charged spherical shell of massless scalar fields numerically.
Crisostomo and Olea (2004) used Hamiltonian treatment to study
gravitational collapse of thin shells. Recently, Sharif and Abbas
(2012) explored the dynamics of scalar field charged thin-shell and
concluded that for both (massless and massive scalar fields) shell
can expand to infinity or collapse to zero size forming a curvature
singularity or bounce under suitable conditions. Many interesting
results regarding thin-shell formalism can be found in literature
(Sharif and Ahmad 2008; Sharif and Iqbal 2009; Sharif and Abbas
2011).

It is well-known that BH solutions (Schwarzschild,
Reissner-Nordstr$\ddot{o}$m (RN), Kerr and Kerr-Newman) contain
curvature singularity beyond their event horizons. A comprehensive
understanding of BH requires singularity-free solutions. Black holes
having regular centers are called regular or nonsingular BHs.
Regular BHs are static and asymptotically flat, satisfying the weak
energy condition. These are exact solutions of the Einstein field
equations for which singularity is avoided in the presence of
horizons (the exterior Schwarzschild-like horizon and an interior
de-Sitter-like horizon). The first regular BH was introduced by
Bardeen (1968) which has both event (exterior) and Cauchy (interior)
horizons, but with a regular center. Bardeen BH can be interpreted
as a magnetic solution of the Einstein equations coupled to
nonlinear electrodynamics where singularity is replaced by a de
Sitter core. Later, many spherically symmetric regular BH solutions
were found based on Bardeen's proposal (Borde 1997; Ayon-Beato and
Garcia 1998; Bronnikov 2000, 2001). Further analysis of singularity
avoidance has been proposed by Hayward (2006). This BH consists of a
compact spacetime region of trapped surfaces, with inner and outer
boundaries which join circularly as a single smooth trapping horizon

The purpose of this paper is to study the dynamical effects of
scalar field on magnetically charged thin-shell using Israel
thin-shell formalism for a class of regular BHs. The format of the
paper is as follows. In the next section, we derive equation of
motion for the thin-shell using Israel formalism. Section \textbf{3}
investigates the equation of motion for a class of regular BHs for
both massless and massive scalar fields. Final remarks are given in
the last section.

\section{Thin-Shell Formalism and Equation of Motion}

Thin-shell formalism (Israel 1966, 1967) has extensively been used
to study the dynamics of matter fields, wormholes, collision of
shells, interior structure of BHs, bubble dynamics and inflationary
scenarios. In this method, surface properties are described in terms
of jump of the extrinsic curvature (functions of intrinsic
coordinates of the layer) across the boundary layer. This formalism
allows to choose four-dimensional coordinates independently on both
sides of the boundary layer. The governing equations resulting from
this formalism correspond to the equation of motion whose solution
can completely describe the dynamics of the shell.

We assume three-dimensional timelike boundary surface $\Sigma$,
which splits spherically symmetric spacetime into two
four-dimensional manifolds $N^{+}$ and $N^{-}$. The interior and
exterior regions are described by a metric of the form
\begin{equation}\label{1}
ds^{2}=F_{\pm}(R)dT^{2}-F^{-1}_{\pm}(R)dR^{2}-R^{2}(d\theta^{2}+\sin^{2}\theta
d\varphi^{2}),
\end{equation}
where explicit forms of $F(R)$ distinguish between different
spacetime geometries. In this paper, we shall use the following
choices of this function.\\\\
\textbf{Table 1:} Regular BHs.
\begin{table}[bht]
\centering
\begin{small}
\begin{tabular}{|c|c|}
\hline\textbf{Name of Regular BH}&\textbf{$F(R)$}\\
\hline\textbf{Bardeen}&$F_{\pm1}(R)=1-\frac{2M_{\pm}R^{2}}{(R^{2}+e_{\pm}^{2})^\frac{3}{2}}$\\
\hline\textbf{Hayward}&$F_{\pm2}(R)=1-\frac{2M_{\pm}R^{2}}{R^{3}+2e_{\pm}^{2}}$\\
\hline\textbf{ABGB}&$F_{\pm3}(R)=1-
\frac{2M_{\pm}R^{2}}{(R^{2}+e_{\pm}^{2})^\frac{3}{2}}+
\frac{e_{\pm}^{2}R^{2}}{(R^{2}+e_{\pm}^{2})^{2}}$\\
\hline
\end{tabular}
\end{small}
\end{table}\\
where $M_{\pm}$ and $e_{\pm}$ are the mass and monopole charge of a
self-gravitating magnetic field of a non-linear electrodynamics
source, respectively. All the above BH solutions correspond to the
Schwarzschild BH for $e=0$. Moreover, it is assumed that the
interior region contains more mass than the exterior region, i.e.,
gravitational masses are unequal $M_{-}\neq M_{+}$ while charge is
uniformly distributed in both regions, i.e., $e=e_{-}=e_{+}$. By
applying the intrinsic coordinates $(\tau,~\theta,~\varphi)$ on the
hypersurface $(\Sigma)$ at $R=R(\tau)$, Eq.(\ref{1}) becomes
\begin{equation}\label{2}
(ds)_{\pm\Sigma}^{2}=\left[F_{\pm}(R)-F_{\pm}^{-1}(R)(\frac{dR}{d\tau})^{2}
(\frac{d\tau}{dT})^{2}\right]dT^{2}-R^{2}(\tau)(d\theta^{2}+\sin^{2}\theta
d\varphi^{2}).
\end{equation}
Here, it is assumed that $T(\tau)$ is a timelike coordinate, i.e.,
$g_{00}>0$. Also, the induced metric on the boundary surface is
given as
\begin{equation}\label{3}
(ds)^{2}=d\tau^{2}-\alpha^{2}(\tau)(d\theta^{2}+\sin^{2}\theta
d\varphi^{2}).
\end{equation}

The continuity of first fundamental forms give
\begin{equation}\label{4}
\left[F_{\pm}(R)-F_{\pm}^{-1}(R)(\frac{dR}{d\tau})^{2}
(\frac{d\tau}{dT})^{2}\right]^{\frac{1}{2}}dT=(d\tau)_{\Sigma},
\quad R(\tau)=\alpha(\tau)_{\Sigma}.
\end{equation}
The outward unit normals $\eta^{\pm}_{\mu}$ in $N^{\pm}$ coordinates
are calculated as
\begin{equation}\nonumber
\eta^{\pm}_{\mu}=(-\dot{R}, \dot{T}, 0, 0),
\end{equation}
where dot represents differentiation with respect to $\tau$. The
surface stress energy-momentum tensor is defined as
\begin{equation}\label{6}
S_{\mu\nu}=\frac{1}{\kappa}([K_{\mu\nu}]-\gamma_{\mu\nu}[K]),
\end{equation}
where $\gamma_{\mu\nu}$ denotes the induced metric, $\kappa$ is the
coupling constant and
\begin{equation}\label{0}
[K_{\mu\nu}]=K^{+}_{\mu\nu}-K^{-}_{\mu\nu},\quad
[K]=\gamma^{\mu\nu}[K_{\mu\nu}].
\end{equation}
The non-vanishing components of extrinsic curvature are
\begin{equation}\label{5}
K^{\pm}_{\tau\tau}=\frac{d}{dR}\sqrt{\dot{R}^{2}+F_{\pm}}, \quad
K^{\pm}_{\theta\theta}=-R\sqrt{\dot{R}^{2}+F_{\pm}}, \quad
K^{\pm}_{\varphi\varphi}=\sin^{2}\theta K^{\pm}_{\theta\theta}.
\end{equation}
The surface stress energy-momentum tensor for a perfect fluid is
\begin{equation}\label{7}
S_{\mu\nu}=(\rho+p)u_{\mu}u_{\nu}-p\gamma_{\mu\nu},
\end{equation}
where $\rho$ is the energy density, $p$ is the isotropic pressure
and $u_{\mu}=\delta^{0}_{\mu}$ is the velocity of the shell. Using
Eqs.(\ref{4}), (\ref{6}) and (\ref{7}), we find
\begin{equation}\label{8}
\rho=\frac{2}{\kappa R^{2}}[K_{\theta\theta}],\quad
p=\frac{1}{\kappa}([K_{\tau\tau}]-\frac{[K_{\theta\theta}]}{R^{2}}).
\end{equation}
Using Eq.(\ref{5}) in (\ref{8}), we can obtain the following
relations
\begin{eqnarray}\label{9}
(\omega_{+}-\omega_{-})+\frac{\kappa}{2}\rho R=0,\\
\label{10}\frac{d}{dR}(\omega_{+}-\omega_{-})+
\frac{1}{R}(\omega_{+}-\omega_{-})-\kappa p=0,
\end{eqnarray}
where $\omega_{\pm}=\sqrt{\dot{R}^{2}+F_{\pm}}$.

The above equations lead to the following differential equation
\begin{equation}\label{11}
\frac{d\rho}{dR}+\frac{2}{R}(\rho+p)=0,
\end{equation}
which is equivalent to the energy conservation of the thin-shell
\begin{equation}\label{12}
\dot{m}+p\dot{A}=0,
\end{equation}
where $m=\rho A$ and $A=4\pi R^{2}(\tau)$ represent mass and area of
the shell, respectively. It is mentioned here that Eq.(\ref{11}) can
be solved using the equation of state $p=\check{k}\rho$ whose
solution is
\begin{equation}\label{13}
\rho=\rho_{0}\left(\frac{R_{0}}{R}\right)^{2(\check{k}+1)},
\end{equation}
where $R_{0}$ represents initial position of the shell at time
$\tau=\tau_{0}$ and $\rho_{0}$ denotes matter density of the shell
at $R_{0}$. Using the above equation, the mass of the shell takes
the form
\begin{equation}\label{14}
m=4\pi\rho_{0}\frac{R_{0}^{2(\check{k}+1)}}{R^{2\check{k}}}.
\end{equation}
Equation (\ref{9}) leads to the equation of motion of the shell
\begin{equation}\label{15}
\dot{R}^{2}+V_{eff}=0,
\end{equation}
where
\begin{equation}\label{16}
V_{eff}(R)=\frac{1}{2}(F_{+}+F_{-})-\frac{(F_{+}+F_{-})^{2}}{(\kappa
\rho R)^{2}}-\frac{1}{16}(\kappa \rho R)^{2},
\end{equation}
is the effective potential which describes shell's motion.

\section{Analysis of Equation of Motion}

Here we study the dynamical behavior of the scalar shell for a
family of regular BHs. For this purpose, we first calculate the
effective potential and the corresponding velocity of the shell with
respect to the stationary observer. We investigate the effect of
charge parameter on the dynamics of the shell.

\subsection{Bardeen Regular BH}

Bardeen (1968) found a solution of the field equations having an
event horizon but excluding a singularity at the origin. This
solution is parameterized by mass $M$ and monopole charge  $e$ of a
self-gravitating magnetic field described by nonlinear
electrodynamics which is well-defined for $R\geq0$, behaves like de
Sitter black hole for $R\rightarrow 0$ and asymptotically as the RN
solution. The corresponding effective potential is
\begin{equation}\label{17}
V_{eff_{(1)}}(R)=1-\left(\frac{2R^{3}}{(R^{2}+
e^{2})^{\frac{3}{2}}}\right)^{2}\left(\frac{M_{+}-M_{-}}{m}\right)^{2}
-\frac{(M_{+}+M_{-})R^{2}}{(R^{2}+e^{2})^{\frac{3}{2}}}-\left(\frac{m}{2R}\right)^{2}.
\end{equation}
To examine the effect of charge on the shell's dynamics,
Eq.(\ref{15}) can be written as
\begin{equation}\label{18}
\dot{R}=\pm\left[\left(\frac{2R^{3}}{(R^{2}+
e^{2})^{\frac{3}{2}}}\right)^{2}\frac{(M_{+}-M_{-})^{2}}{m^{2}}
+\frac{(M_{+}+M_{-})R^{2}}{(R^{2}+e^{2})^{\frac{3}{2}}}+\left(\frac{m}{2R}\right)^{2}
-1\right]^{\frac{1}{2}}.
\end{equation}
Here $-(+)$ correspond to collapse (expansion) of the shell. In
Figure \textbf{1}, the left graph represents $\dot{R}>0$ while the
right graph shows the behavior of shell's velocity as $\dot{R}<0$.
In the first case, the velocity of the shell decreases positively
while in the second case, it increases negatively. We also see that
velocity of the charged (regular) shell is less than the uncharged
(singular) in both cases and also both curves match for the large
radius.
\begin{figure}\centering
\epsfig{file=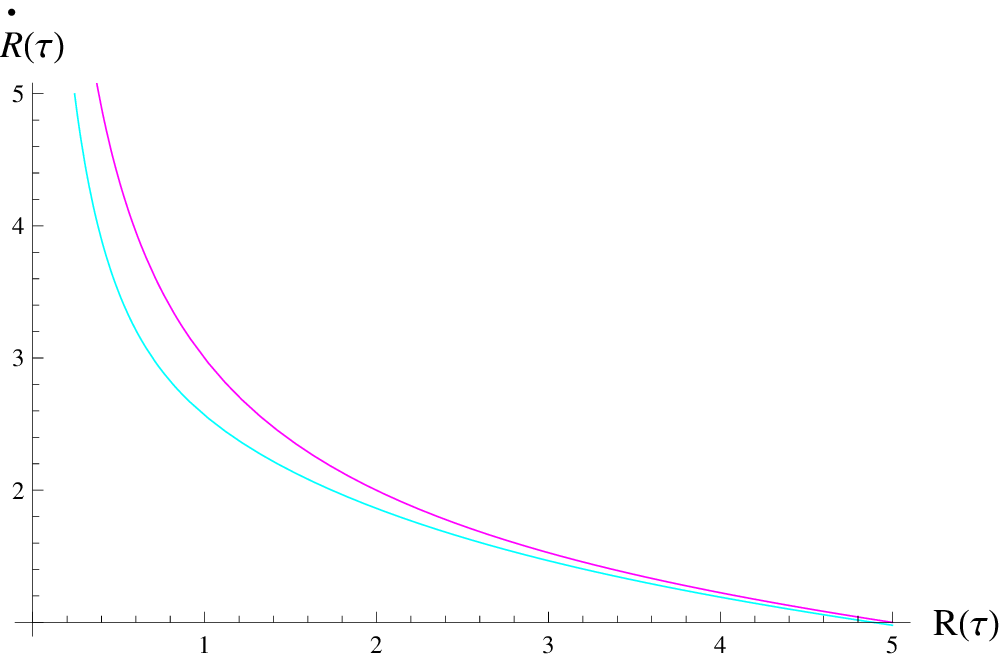,width=.42\linewidth}\epsfig{file=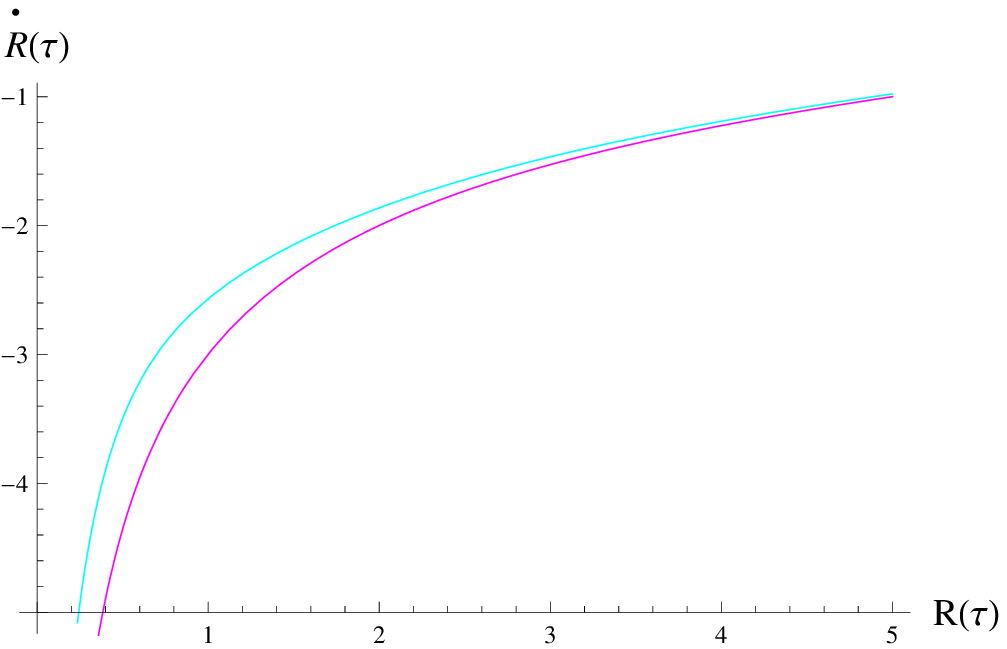,width=.42\linewidth}
\caption{Plots of $\dot{R}({\tau})$ versus $R$ for Bardeen BH  when
$M_{-}=0,~M_{+}=1$, $R_{0}=\rho_{0}=\check{k}=1$ and $e=1$. Here
$M_{+}=1$ corresponds to mass of the exterior BH while $M_{-}=0$
denotes the interior flat geometry. Left and right graphs describe
the case of expansion and collapse while blue and purple curves
correspond to charge and uncharged shell.}
\end{figure}

\subsection{Hayward BH}

Hayward (2006) found a simple regular BH solution in which $e$ is
related to the cosmological constant $\Lambda$ as
$e^{2}=\frac{3M}{\Lambda}$ and for the well-defined asymptotic
limits, this corresponds to the Schwarzschild BH as
$R\rightarrow\infty$ while it becomes de Sitter spacetime at the
center ($R\rightarrow0$). The corresponding effective potential is
\begin{equation}\label{tm}
V_{eff_{(2)}}(R)=1-\left(\frac{2R^{3}}{R^{3}+
2e^{2}}\right)^{2}\left(\frac{M_{+}-M_{-}}{m}\right)^{2}
-\frac{(M_{+}+M_{-})R^{2}}{(R^{3}+
2e^{2})}-\left(\frac{m}{2R}\right)^{2}.
\end{equation}
Equation (\ref{15}) and (\ref{tm}) yield
\begin{equation}\label{tt}
\dot{R}=\pm\left[\left(\frac{2R^{3}}{R^{3}+
2e^{2}}\right)^{2}\left(\frac{M_{+}-M_{-}}{m}\right)^{2}
+\frac{(M_{+}+M_{-})R^{2}}{R^{3}+
2e^{2}}+\left(\frac{m}{2R}\right)^{2} -1\right]^{\frac{1}{2}}.
\end{equation}
Figure \textbf{2} shows the same behavior of velocity as for Bardeen
regular BH.
\begin{figure}\centering
\epsfig{file=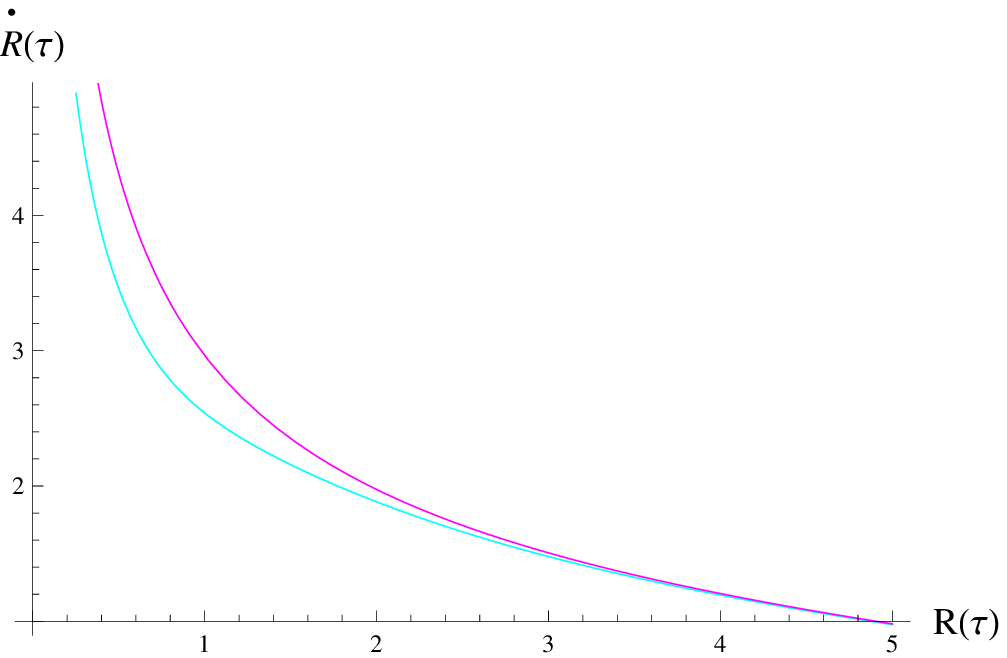,width=.42\linewidth}\epsfig{file=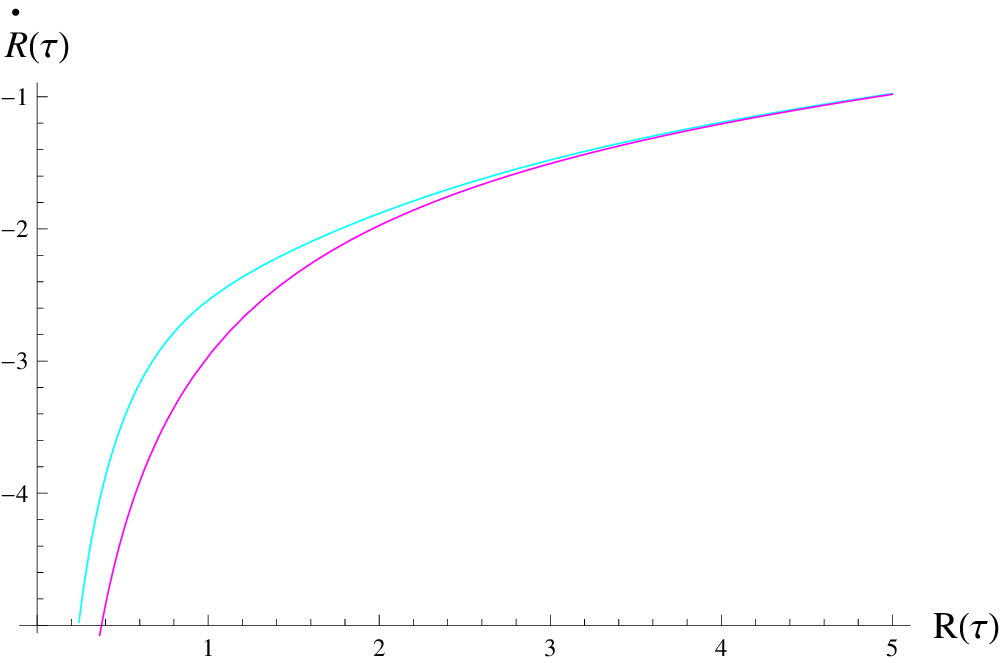,width=.42\linewidth}
\caption{Plots of $\dot{R}({\tau})$ versus $R$ for Hayward BH.}
\end{figure}

\subsection{Ayon-Beato and Garcia and Bronnikov BH}

Another interesting model of non-linear electrodynamics coupled with
GR was proposed by Ayon-Beato and Garcia (1998) and Bronnikov (2000,
2001) (ABGB). This is singularity free BH for a nonlinear magnetic
monopole which behaves asymptotically as RN solution. The
corresponding effective potential is
\begin{eqnarray}\nonumber
V_{eff_{(3)}}(R)&=&1-\left(\frac{2R^{3}}{(R^{2}+
e^{2})^{\frac{3}{2}}}\right)^{2}\left(\frac{M_{+}-M_{-}}{m}\right)^{2}
-\frac{(M_{+}+M_{-})R^{2}}{(R^{2}+e^{2})^{\frac{3}{2}}}\\\label{tttn}&+&
\frac{e^{2}R^{2}}{(R^{2}+e^{2})^{2}}-\left(\frac{m}{2R}\right)^{2}.
\end{eqnarray}
Solving Eqs.(\ref{15}) and (\ref{tttn}), we obtain
\begin{eqnarray}\nonumber
\dot{R}&=&\pm\left[\left(\frac{2R^{3}}{(R^{2}+
e^{2})^{\frac{3}{2}}}\right)^{2}\frac{(M_{+}-M_{-})^{2}}{m^{2}}
+\frac{(M_{+}+M_{-})R^{2}}{(R^{2}+e^{2})^{\frac{3}{2}}}
\right.\\\label{t3}&-&\left.\frac{e^{2}R^{2}}{(R^{2}+e^{2})^{2}}
+\left(\frac{m}{2R}\right)^{2} -1\right]^{\frac{1}{2}}.
\end{eqnarray}
This also shows the same behavior as for Bardeen regular BH (Figure:
\textbf{1}).
\begin{figure}\centering
\epsfig{file=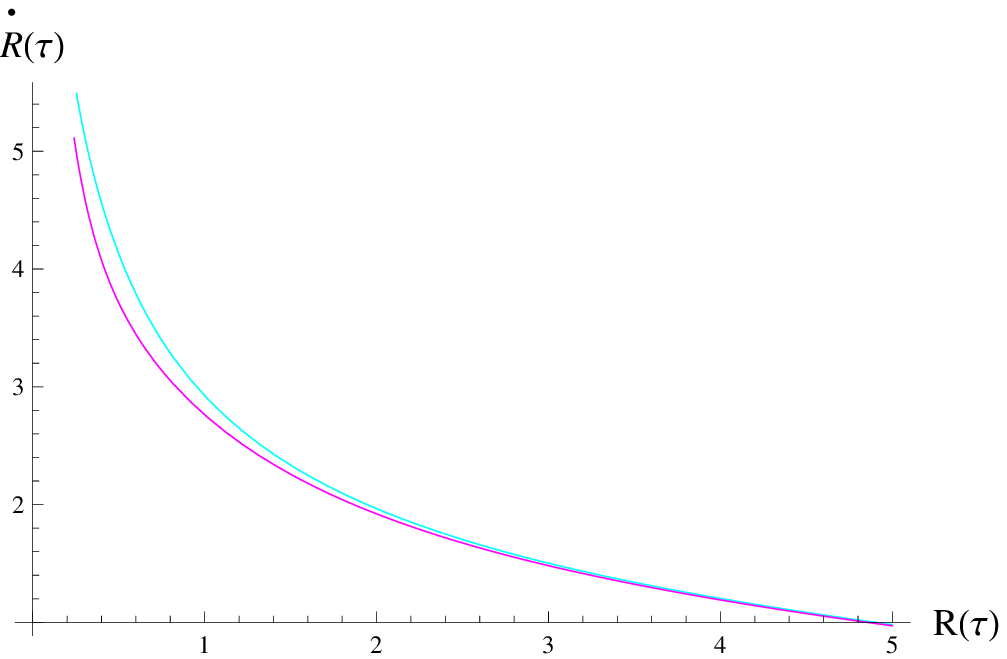,width=.42\linewidth}\epsfig{file=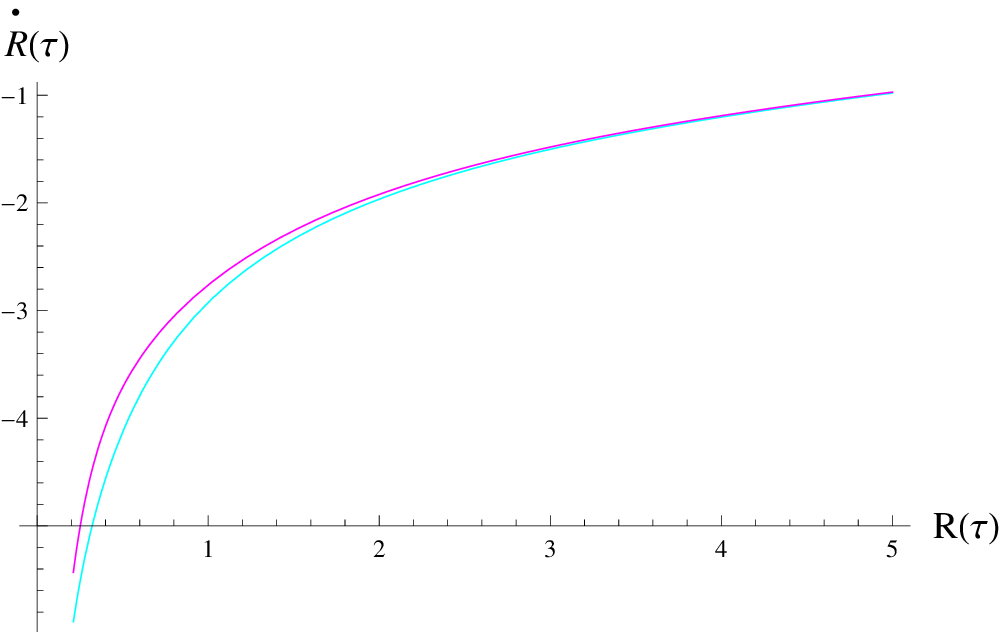,width=.42\linewidth}
\caption{Plots of $\dot{R}({\tau})$ versus $R$ for ABGB BH.}
\end{figure}

\subsection{Dynamics of Scalar Shell}

In this section, we study the dynamics of thin-shell with the scalar
field. For this purpose, we obtain the energy-momentum tensor for
the scalar field by applying a transformation on Eq.(\ref{7}) given
as (N$\acute{u}\tilde{n}$ez 1998)
\begin{equation}\label{19}
u_{\mu}=\frac{\varphi_{,\mu}}{\sqrt{\varphi_{,\nu}\varphi^{,\nu}}},
\quad \rho=\frac{1}{2}[\varphi_{,\nu}\varphi^{,\nu}+2V(\varphi)],
\quad p=\frac{1}{2}[\varphi_{,\nu}\varphi^{,\nu}-2V(\varphi)],
\end{equation}
where $V(\varphi)=m^{2}\varphi^{2}$ is the potential term
representing a massive scalar field. In the absence of this term,
the scalar field will become massless. From Eqs.(\ref{7}) and
(\ref{19}), the energy-momentum tensor for the scalar field can be
written as
\begin{equation}\nonumber
S_{\mu\nu}=\nabla_{\mu}\varphi\nabla_{\nu}\varphi-\gamma_{\mu\nu}
[\frac{1}{2}(\nabla\varphi)^{2}-V(\varphi)].
\end{equation}
Since the induced metric is a function of $\tau$ only, so $\varphi$
also depends on $\tau$. Thus Eq.(\ref{19}) takes the form
\begin{equation}\label{20}
\rho=\frac{1}{2}[\dot{\varphi}^{2}+2V(\varphi)], \quad
p=\frac{1}{2}[\dot{\varphi}^{2}-2V(\varphi)].
\end{equation}
The total mass of the scalar shell can be written as
\begin{equation}\label{21}
m=2\pi R^{2}[\dot{\varphi}+2V(\varphi)].
\end{equation}
Inserting Eqs.(\ref{20}) and (\ref{21}) in (\ref{12}), we obtain
\begin{equation}\label{22}
\ddot{\varphi}+\frac{2\dot{R}}{R}\dot{\varphi}+ \frac{\partial
V}{\partial\varphi}=0,
\end{equation}
which is the Klien-Gordon (KG) equation, $\Box\varphi+\frac{\partial
V}{\partial\varphi}=0$, written in shell's coordinate system.

The effective potentials for the Bardeen, Hayward and ABGB BHs in
terms of scalar field are obtained as
\begin{eqnarray}\nonumber
V_{eff_{(1)}}(R)&=&1-\left(\frac{2R^{3}}{(R^{2}+
e^{2})^{\frac{3}{2}}}\right)^{2}\left(\frac{M_{+}-M_{-}}{2\pi
R^{2}(\dot{\varphi}+2V(\varphi))}\right)^{2}
\\\label{23}&-&\frac{(M_{+}+M_{-})R^{2}}{(R^{2}+e^{2})^{\frac{3}{2}}}-[\pi
R(\dot{\varphi}+2V(\varphi))]^{2}, \\\nonumber
V_{eff_{(2)}}(R)&=&1-\left(\frac{2R^{3}}{R^{3}+
2e^{2}}\right)^{2}\left(\frac{M_{+}-M_{-}}{2\pi
R^{2}(\dot{\varphi}+2V(\varphi))}\right)^{2}
\\\label{t1}&-&\frac{(M_{+}+M_{-})R^{2}}{(R^{3}+ 2e^{2})}-[\pi
R(\dot{\varphi}+2V(\varphi))]^{2},
\\\nonumber
V_{eff_{(3)}}(R)&=&1-\left(\frac{2R^{3}}{(R^{2}+
e^{2})^{\frac{3}{2}}}\right)^{2}\left(\frac{M_{+}-M_{-}}{2\pi
R^{2}(\dot{\varphi}+2V(\varphi))}\right)^{2}
\\\label{23t}&-&\frac{(M_{+}+M_{-})R^{2}}{(R^{2}+e^{2})^{\frac{3}{2}}}
+\frac{e^{2}R^{2}}{(R^{2}+e^{2})^{2}}-[\pi
R(\dot{\varphi}+2V(\varphi))]^{2}.
\end{eqnarray}
Now we solve Eqs.(\ref{15}) and (\ref{22}) with the help of
Eqs.(\ref{23})-(\ref{23t}). These equations cannot be solved
analytically, so we solve them numerically, assuming
$M_{-}=0,~M_{+}=1$, $R_{0}=\rho_{0}=\check{k}=1,~e=1$ and $m=1$
which are shown in Figures \textbf{4} and \textbf{5}. The behavior
of shell's radius is represented in Figure \textbf{4} for Bardeen,
Hayward and ABGB BHs. The upper and lower curves represent expanding
and collapsing shell which describe the motion of the shell. For
Bardeen BH, the upper curve starts with a static configuration then
expands forever, while the lower curve represents that the shell
radius starts with a uniform motion then decreases infinitely on the
negative axis. In the plots of radii for Hayward and ABGB BHs, the
upper curves indicate that the shell expands endlessly and the lower
curve shows that the radius is decreasing continuously. Figure
\textbf{5} describes solutions of the KG equation (\ref{22}) for the
same BHs. The upper curves represent collapsing while the lower
curves show the expanding scalar shell. The scalar field density is
increasing in the first case (collapse) while it decays to zero as
time increases in the second case (expansion).
\begin{figure}\centering
\epsfig{file=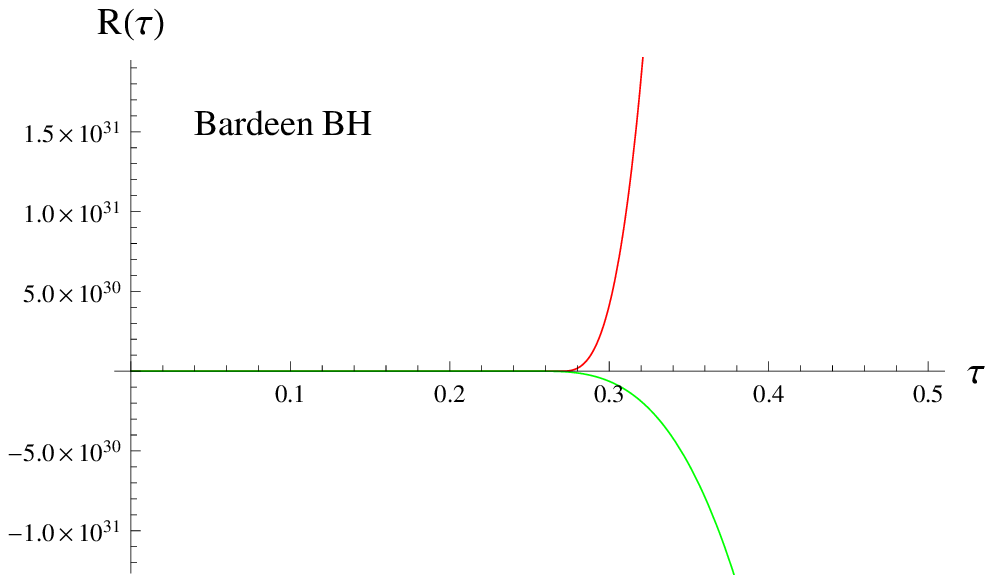,width=.46\linewidth}\epsfig{file=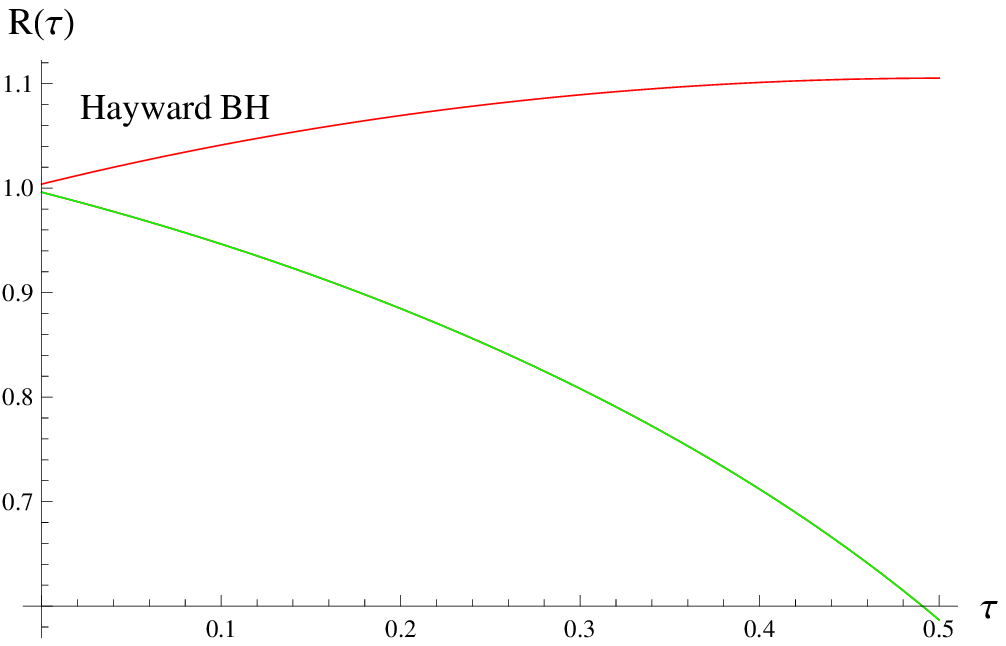,width=.42\linewidth}
\\\epsfig{file=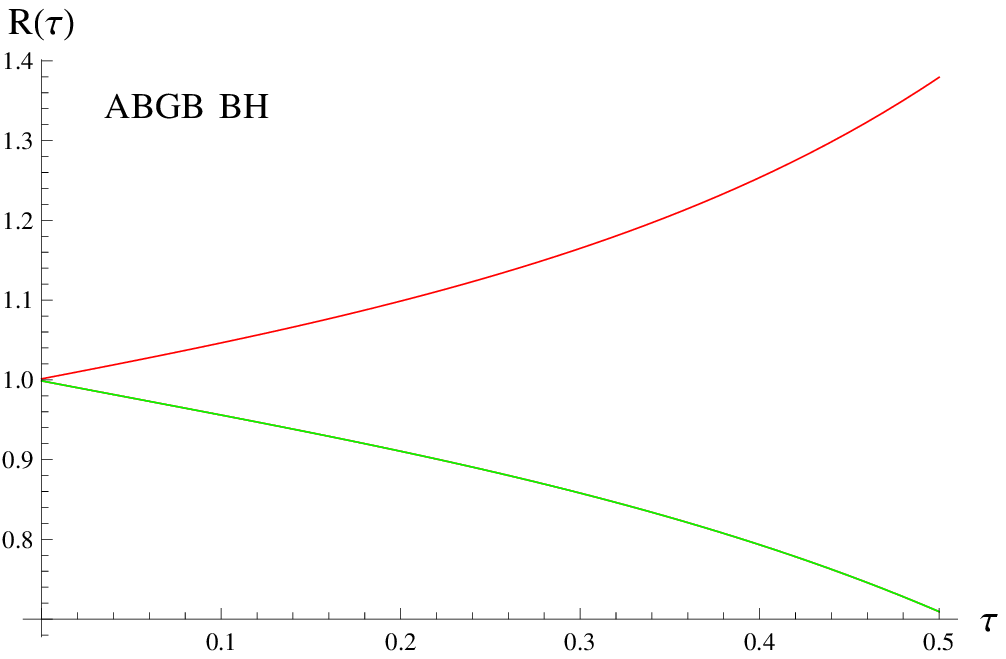,width=.42\linewidth}
\caption{Plots of shell's radius with scalar field.}
\end{figure}
\begin{figure}\centering
\epsfig{file=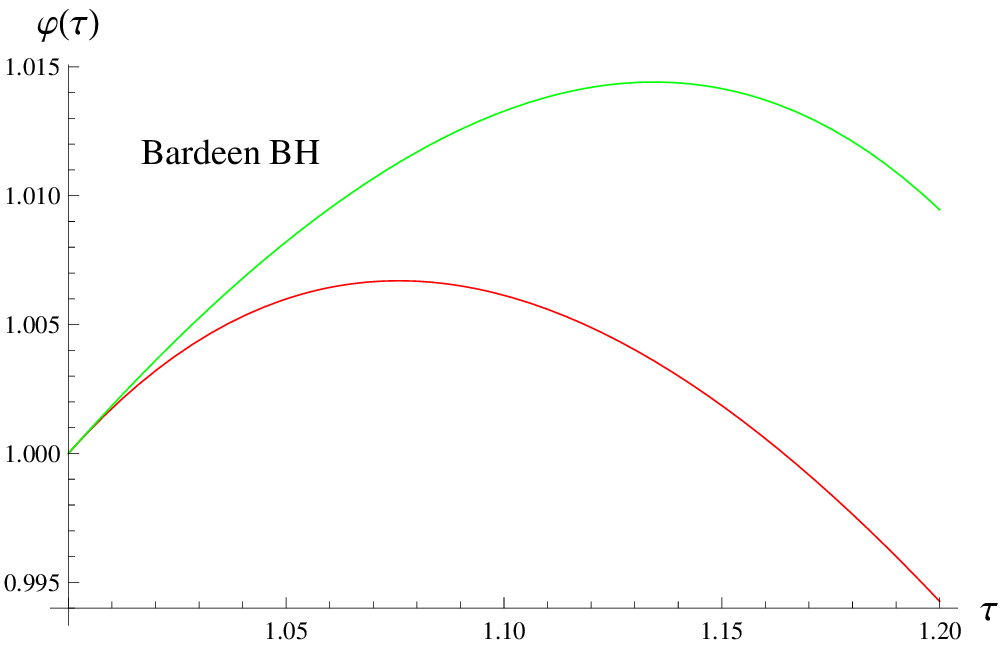,width=.42\linewidth}\epsfig{file=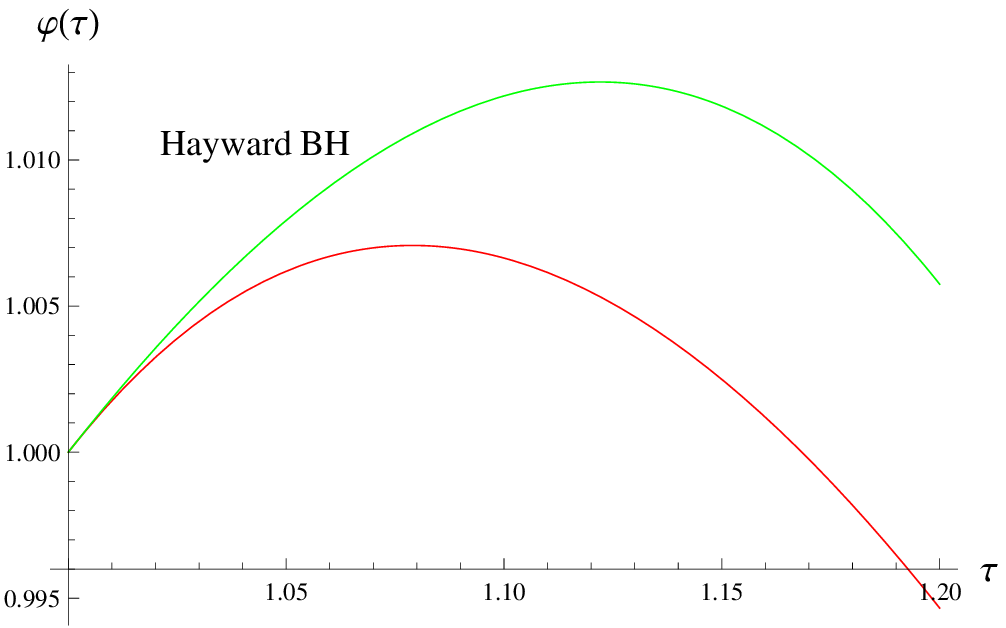,width=.42\linewidth}
\\\epsfig{file=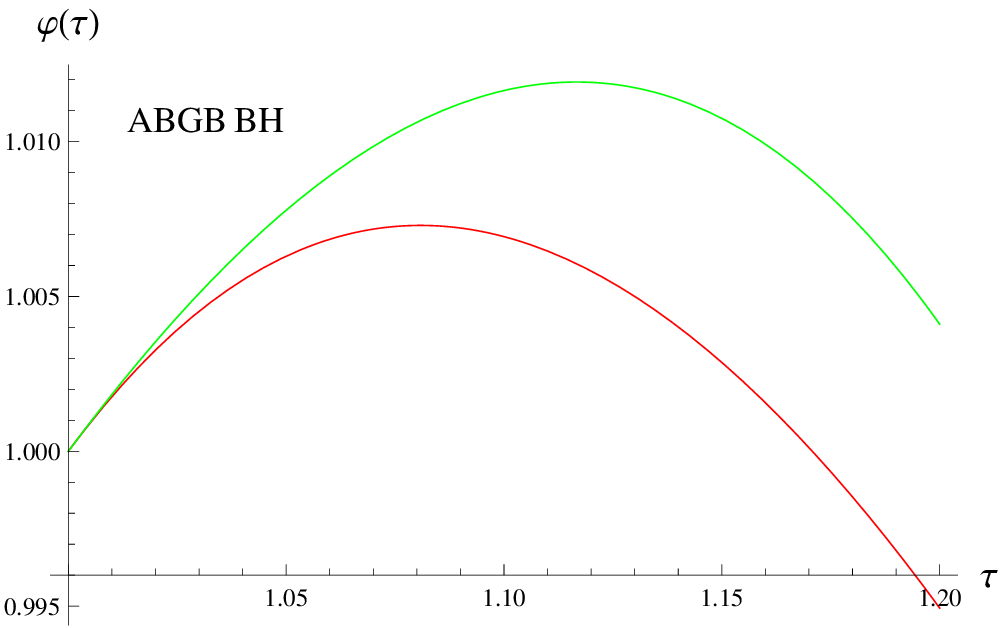,width=.42\linewidth}
\caption{Plots of the scalar field. Green and red curves correspond
to collapse and expansion.}
\end{figure}
\subsubsection{Massless Scalar Shell}

Here we investigate dynamics of the shell in the absence of
$V(\varphi)$, i.e., the massless scalar field. In this case, KG
equation becomes $\ddot{\varphi}+\frac{2\dot{R}}{R}\dot{\varphi}=0$,
and its integration leads to $\dot{\varphi}=\frac{\lambda}{R^{2}}$,
where $\lambda$ is an integration constant. The corresponding
equations of motion for the Bardeen, Hayward and ABGB BHs with
Eq.(\ref{23})-(\ref{23t}) become
\begin{eqnarray}\nonumber
\dot{R}^{2}&+&1-\left(\frac{2R^{5}}{(R^{2}+
e^{2})^{\frac{3}{2}}}\right)^{2}\left(\frac{M_{+}-M_{-}}{2\pi
\lambda^{2}}\right)^{2}
-\frac{(M_{+}+M_{-})R^{2}}{(R^{2}+e^{2})^{\frac{3}{2}}}\\\label{24}&-&
\left(\frac{\pi\lambda^{2}}{R^{3}}\right)^{2}=0,
\\\nonumber
\dot{R}^{2}&+&1-\left(\frac{2R^{5}}{R^{3}+
2e^{2}}\right)^{2}\left(\frac{M_{+}-M_{-}}{2\pi
\lambda^{2}}\right)^{2} -\frac{(M_{+}+M_{-})R^{2}}{R^{3}+
2e^{2}}\\\label{t2}&-&\left(\frac{\pi\lambda^{2}}{R^{3}}\right)^{2}=0,
\\\nonumber
\dot{R}^{2}&+&1-\left(\frac{2R^{5}}{(R^{2}+
e^{2})^{\frac{3}{2}}}\right)^{2}\left(\frac{M_{+}-M_{-}}{2\pi
\lambda^{2}}\right)^{2}
-\frac{(M_{+}+M_{-})R^{2}}{(R^{2}+e^{2})^{\frac{3}{2}}}
\\\label{t4}&+&\frac{e^{2}R^{2}}{(R^{2}+e^{2})^{2}}-
\left(\frac{\pi\lambda^{2}}{R^{3}}\right)^{2}=0.
\end{eqnarray}
These can be simplified by using the parameters
\begin{equation}
[M]=M_{+}-M_{-},\quad \bar{M}=\frac{M_{+}+M_{-}}{2}.
\end{equation}
Inserting these parameters in Eqs.(\ref{24})-(\ref{t4}), we have
\begin{equation}
\dot{R}^{2}+V_{eff}=0,
\end{equation}
where
\begin{eqnarray}\label{25}
V_{eff_{(1)}}(R)&=&1-\left(\frac{2R^{5}}{(R^{2}+
e^{2})^{\frac{3}{2}}}\right)^{2}\left(\frac{[M]}{2\pi
\lambda^{2}}\right)^{2}
-\frac{(2\bar{M})R^{2}}{(R^{2}+e^{2})^{\frac{3}{2}}}-
\left(\frac{\pi\lambda^{2}}{R^{3}}\right)^{2},
\end{eqnarray}
\begin{eqnarray}\label{t2}
V_{eff_{(2)}}(R)&=&1-\left(\frac{2R^{5}}{R^{3}+
2e^{2}}\right)^{2}\left(\frac{[M]}{2\pi \lambda^{2}}\right)^{2}
-\frac{(2\bar{M})R^{2}}{R^{3}+ 2e^{2}}-
\left(\frac{\pi\lambda^{2}}{R^{3}}\right)^{2},
\\\nonumber
V_{eff_{(3)}}(R)&=&1-\left(\frac{2R^{5}}{(R^{2}+
e^{2})^{\frac{3}{2}}}\right)^{2}\left(\frac{[M]}{2\pi
\lambda^{2}}\right)^{2}
-\frac{(2\bar{M})R^{2}}{(R^{2}+e^{2})^{\frac{3}{2}}}
\\\label{t44}&+&\frac{e^{2}R^{2}}{(R^{2}+e^{2})^{2}}-
\left(\frac{\pi\lambda^{2}}{R^{3}}\right)^{2}.
\end{eqnarray}
\begin{figure}\centering
\epsfig{file=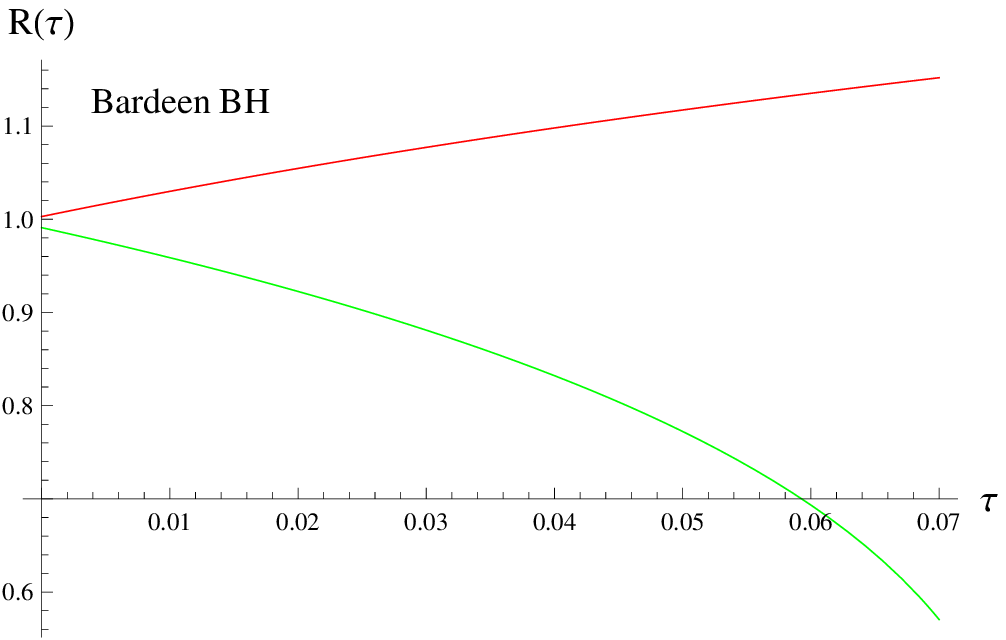,width=.42\linewidth}\epsfig{file=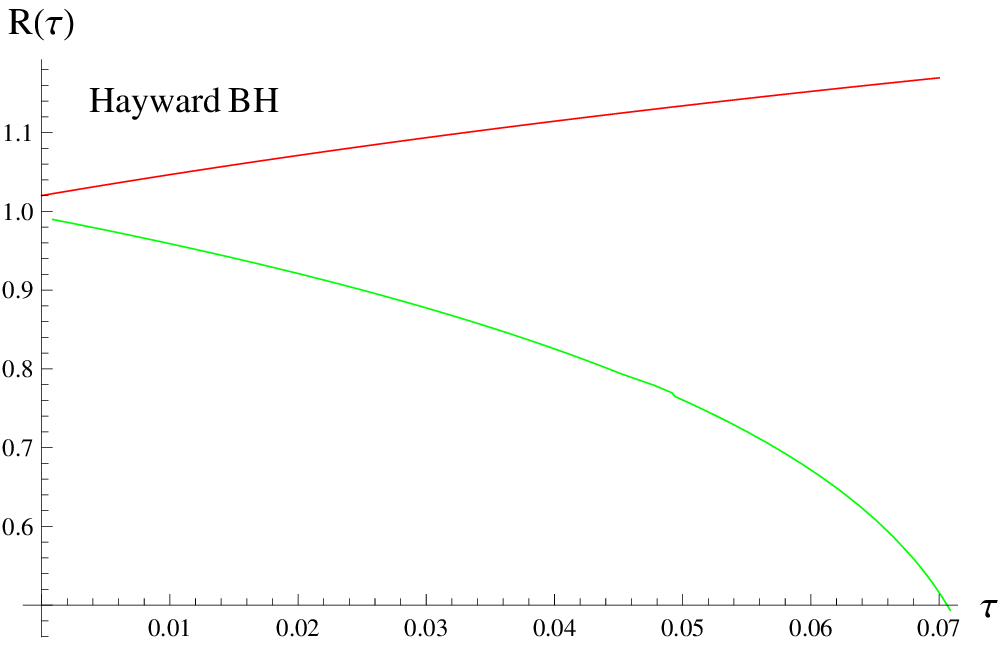,width=.42\linewidth}
\\\epsfig{file=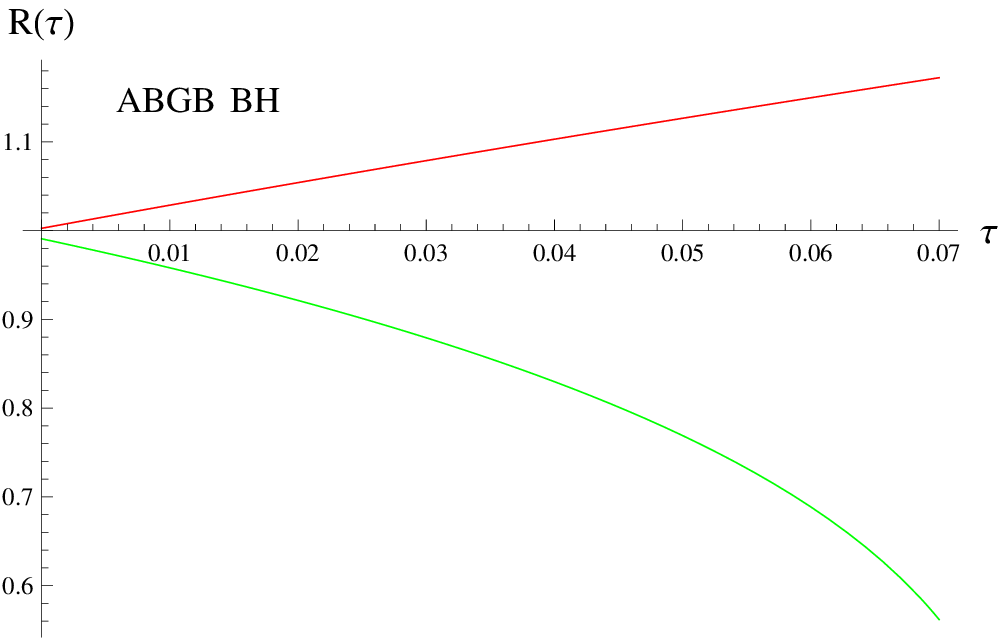,width=.42\linewidth}
\caption{Behavior of radius of the massless scalar shell.}
\end{figure}
\begin{figure}\centering
\epsfig{file=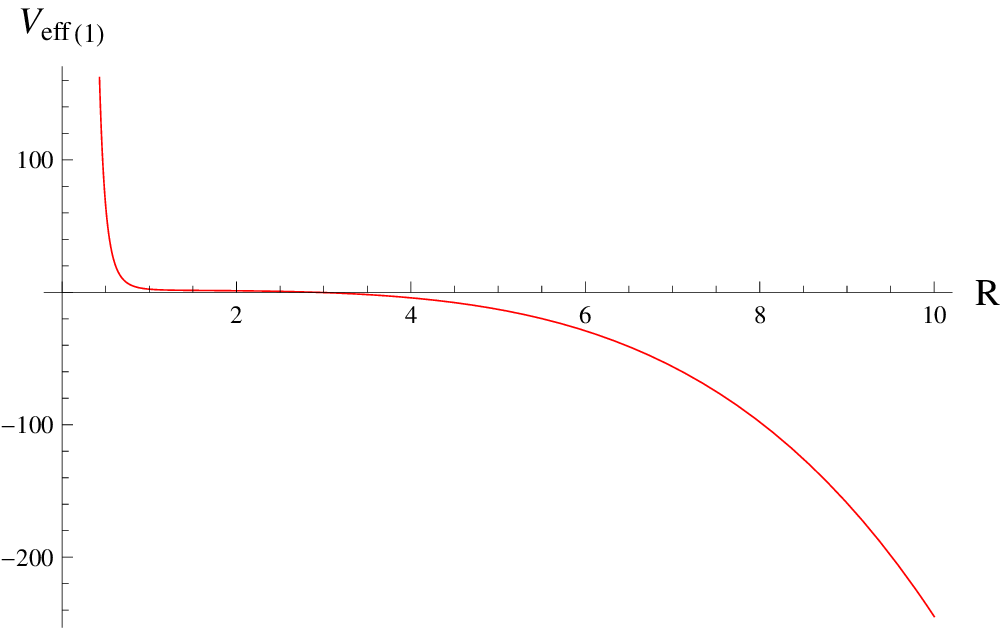,width=.42\linewidth}\epsfig{file=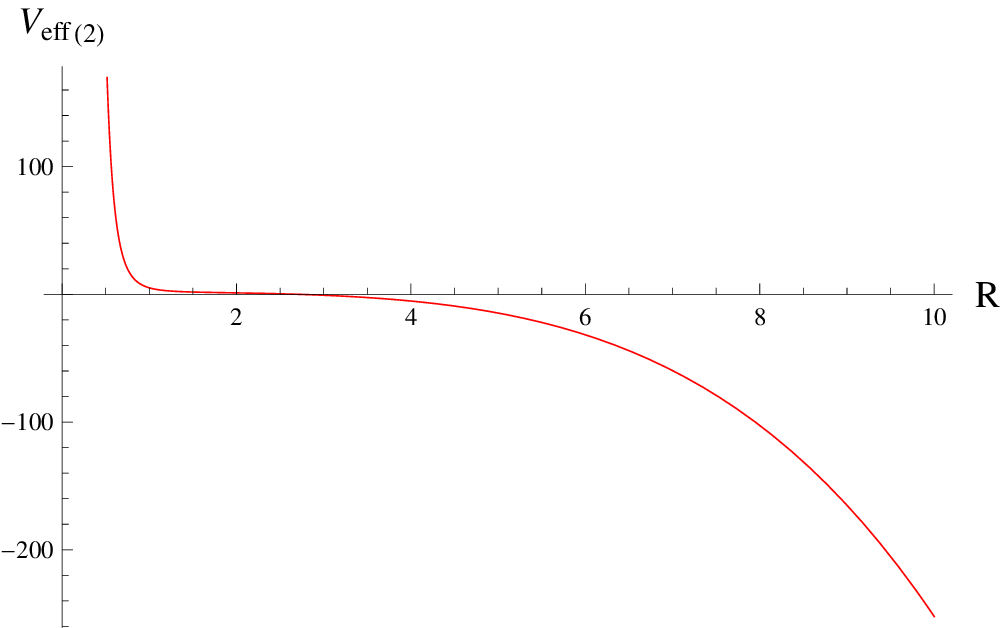,width=.42\linewidth}
\\\epsfig{file=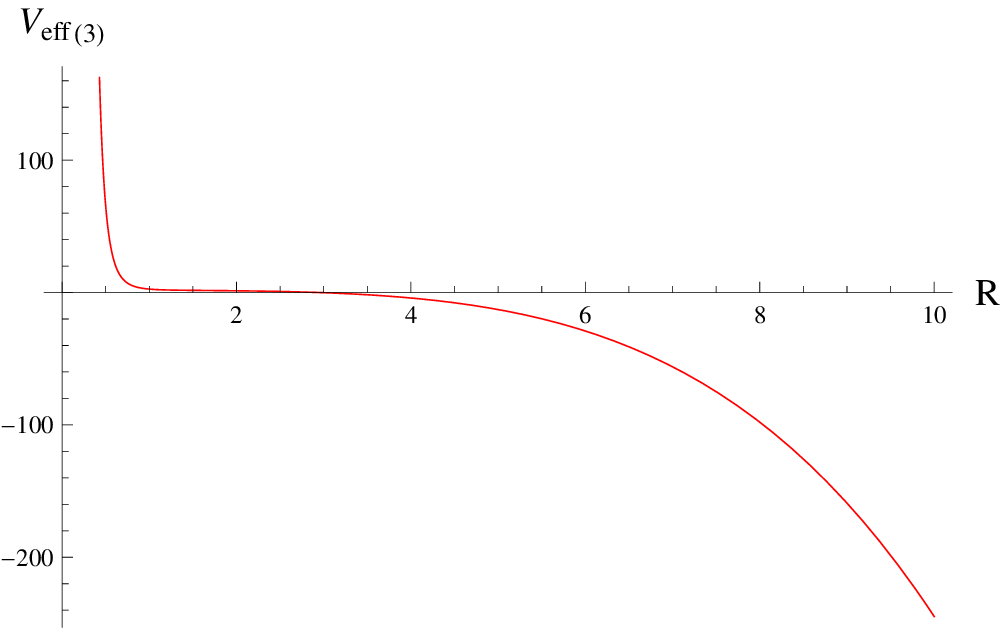,width=.42\linewidth}
\caption{Plots of $V_{eff}$ versus $R$ for the  massless case. In
the upper panel, the left and right graphs show the plots for
Bardeen and Hayward BHs while the lower panel represents ABGB BH for
fixed $M_{-}$ and $M_{+}$.}
\end{figure}
\begin{figure}\centering
\epsfig{file=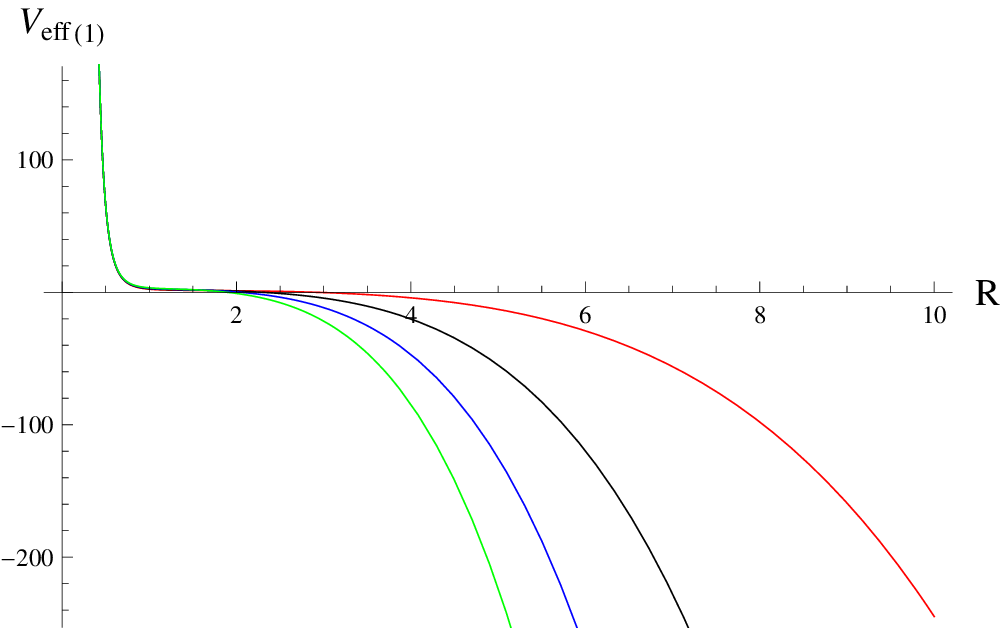,width=.42\linewidth}\epsfig{file=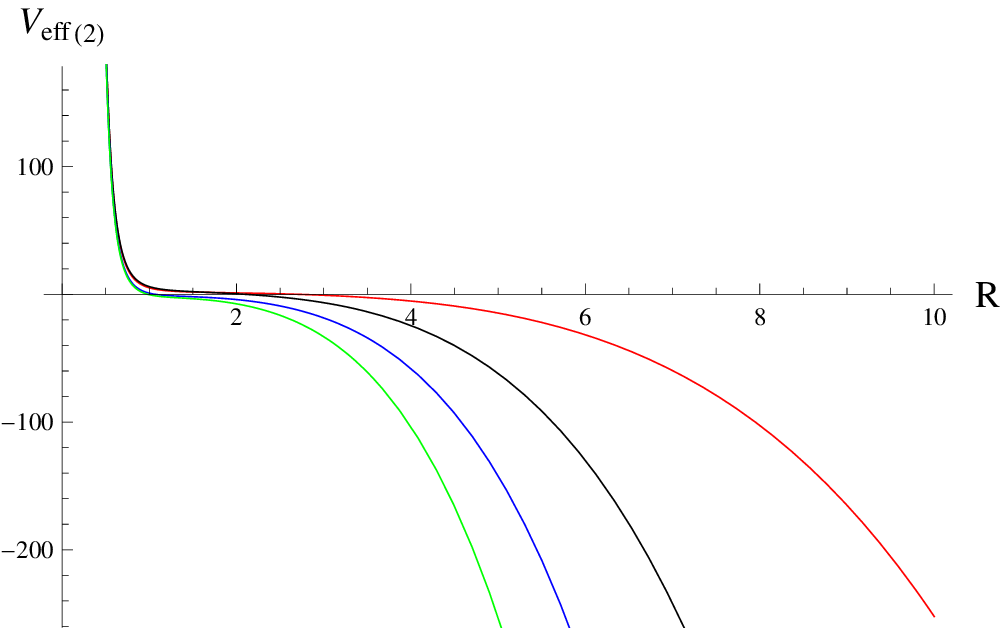,width=.42\linewidth}
\\\epsfig{file=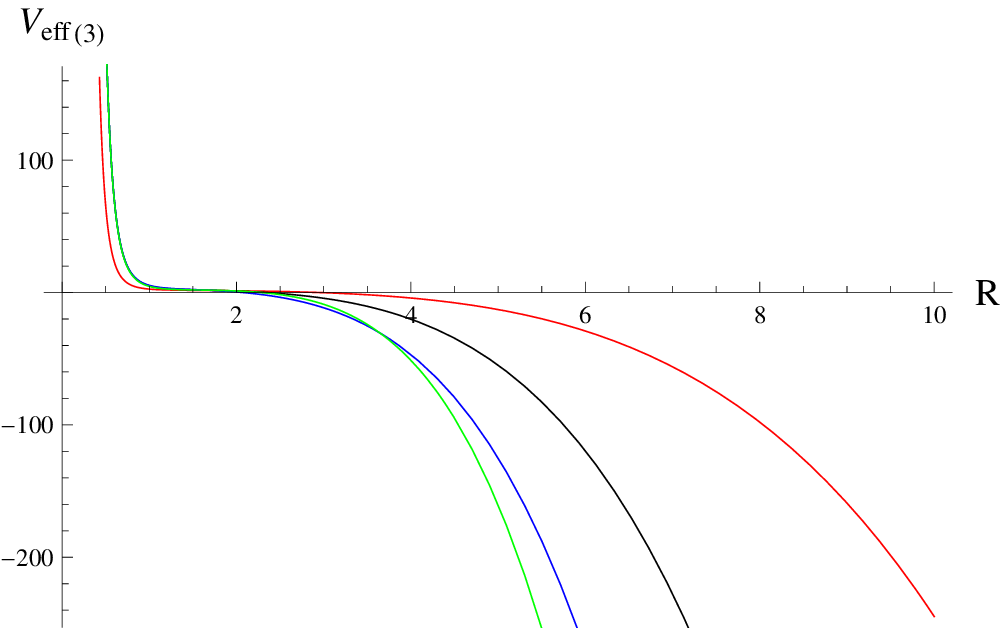,width=.42\linewidth}
\caption{Behavior of $V_{eff}$ of the massless shell by varying
$M_{+}$.}
\end{figure}
\begin{figure}\centering
\epsfig{file=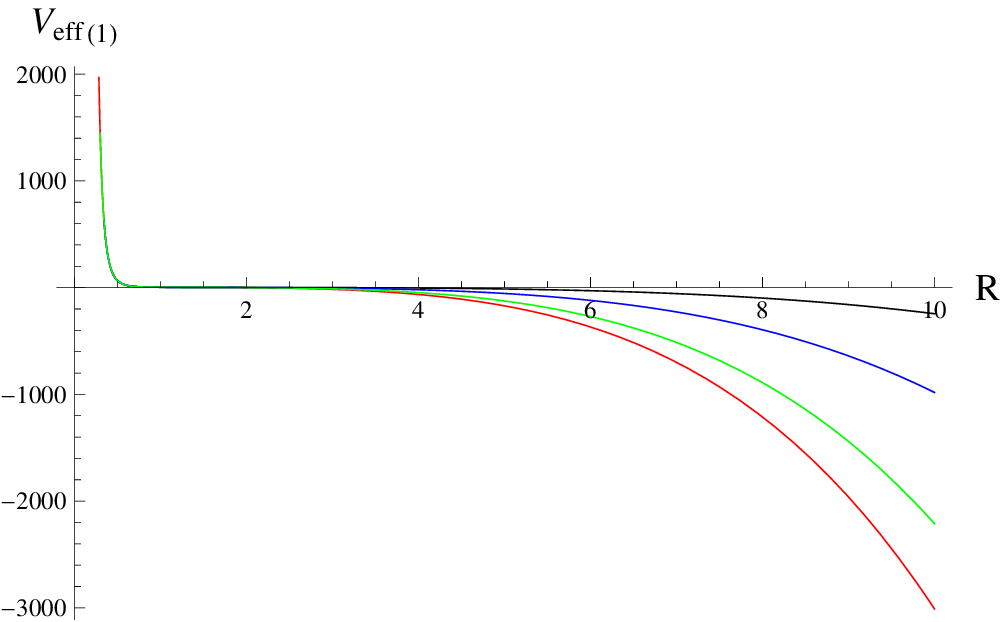,width=.42\linewidth}\epsfig{file=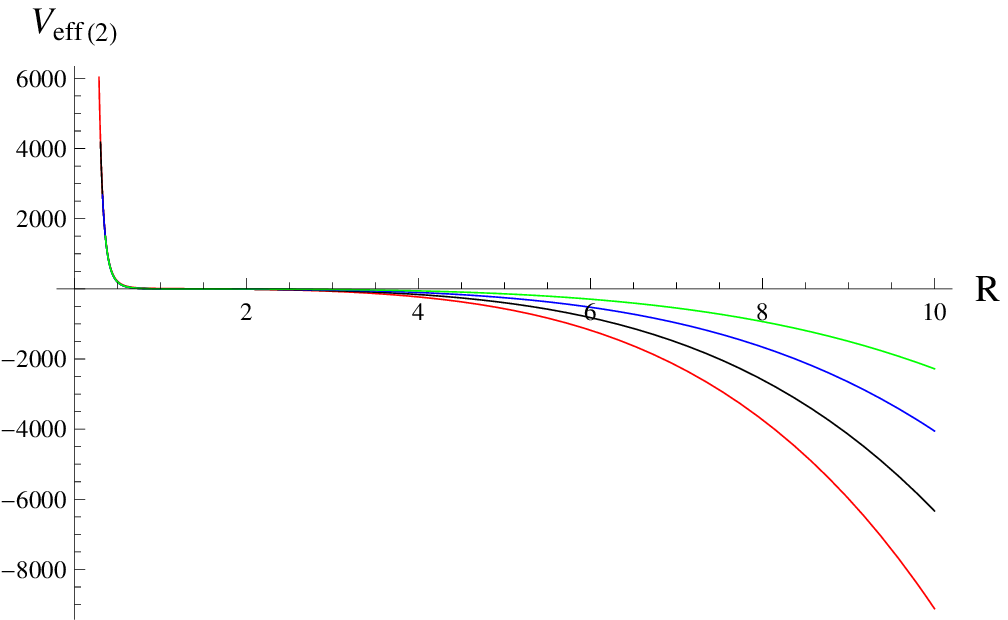,width=.42\linewidth}
\\\epsfig{file=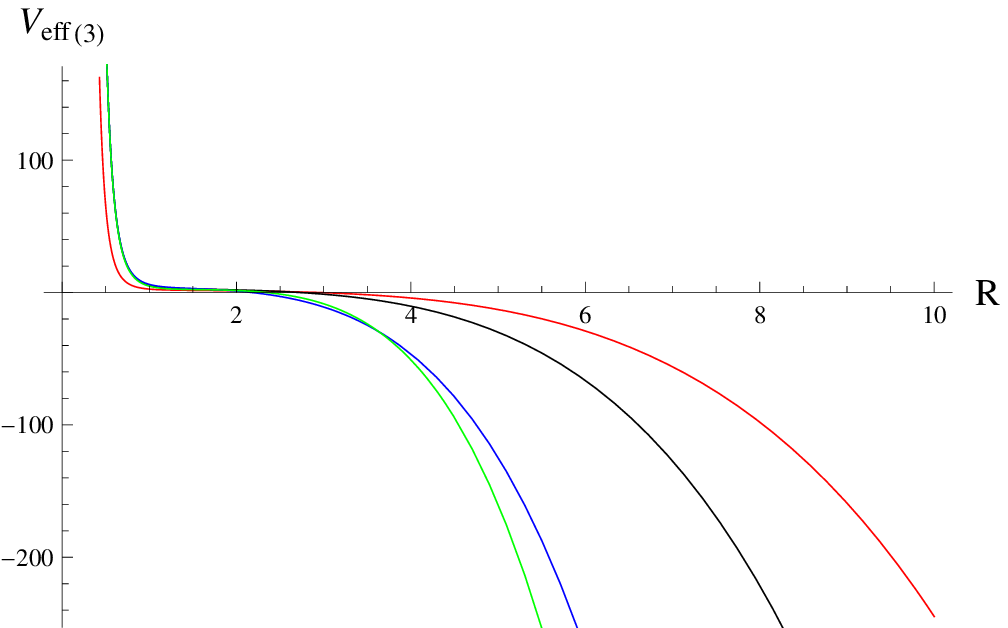,width=.42\linewidth}
\caption{Behavior of $V_{eff}$ of the massless shell by varying
$M_{-}$.}
\end{figure}
\begin{figure}\centering
\epsfig{file=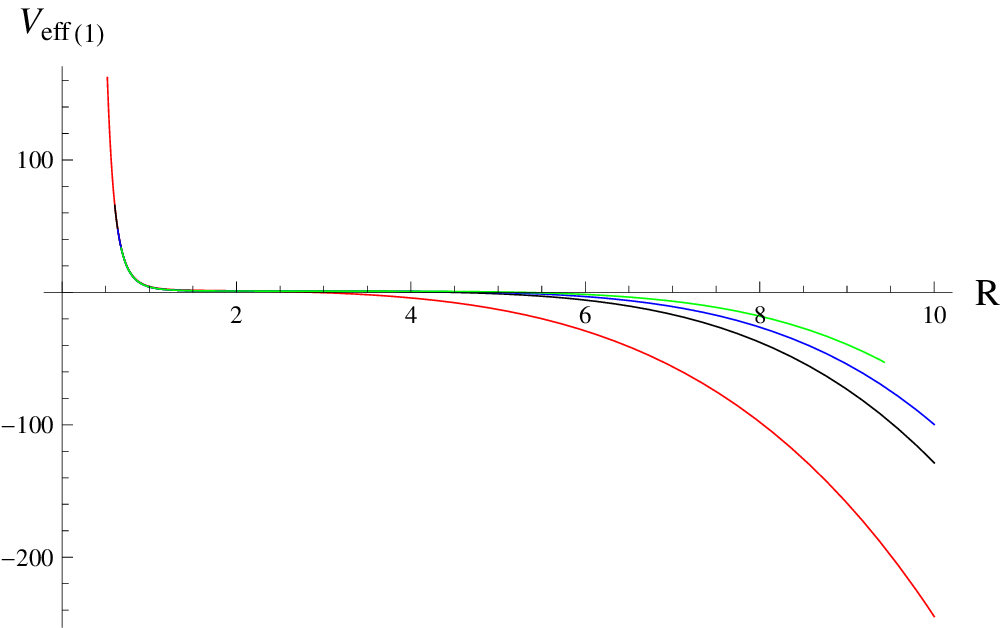,width=.42\linewidth}\epsfig{file=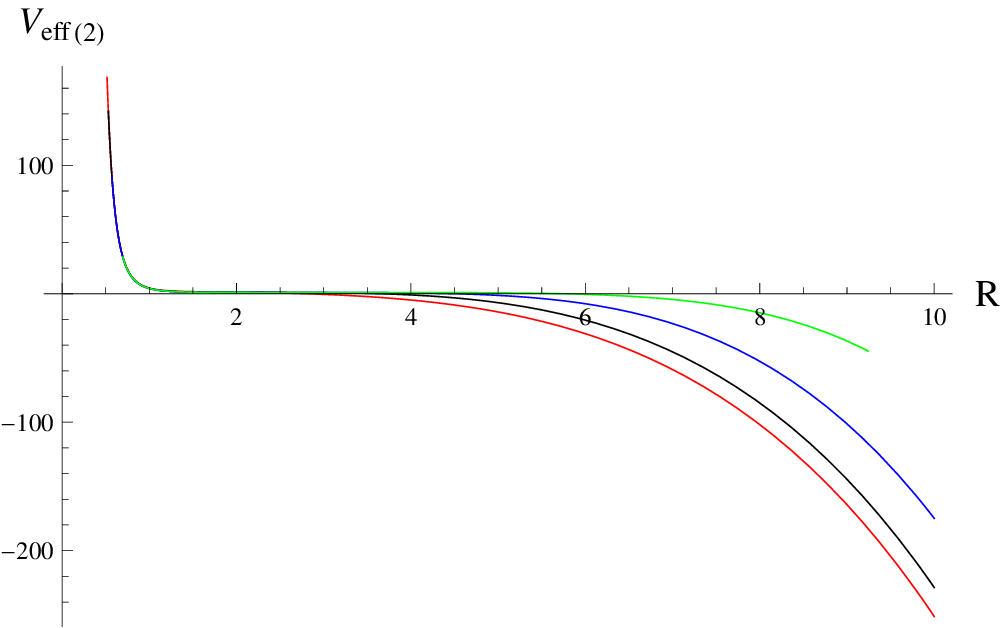,width=.42\linewidth}
\\\epsfig{file=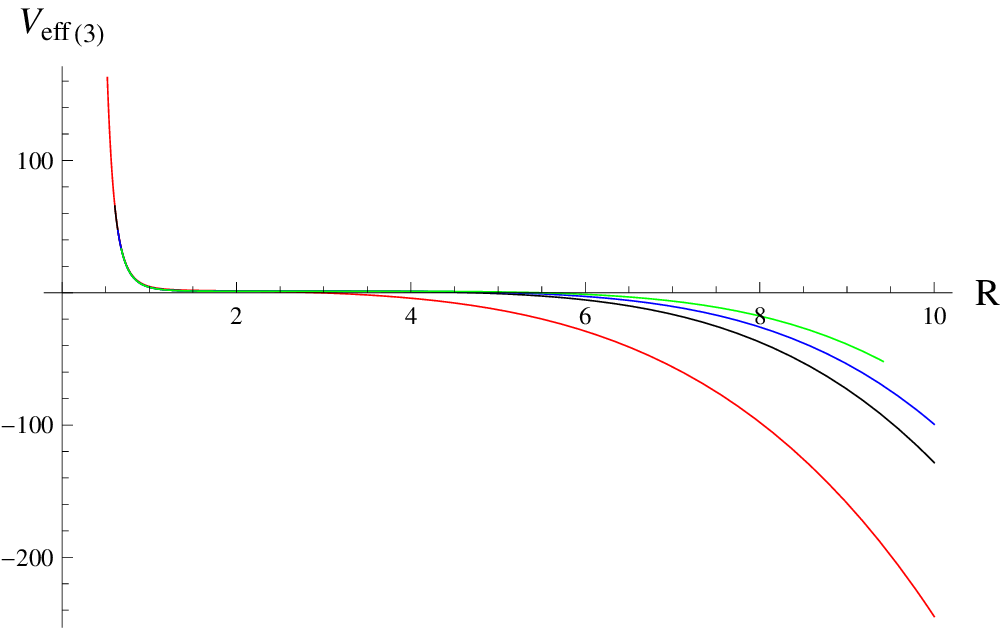,width=.42\linewidth}
\caption{Behavior of $V_{eff}$ of the massless shell by varying
$e$.}
\end{figure}
\begin{figure}\centering
\epsfig{file=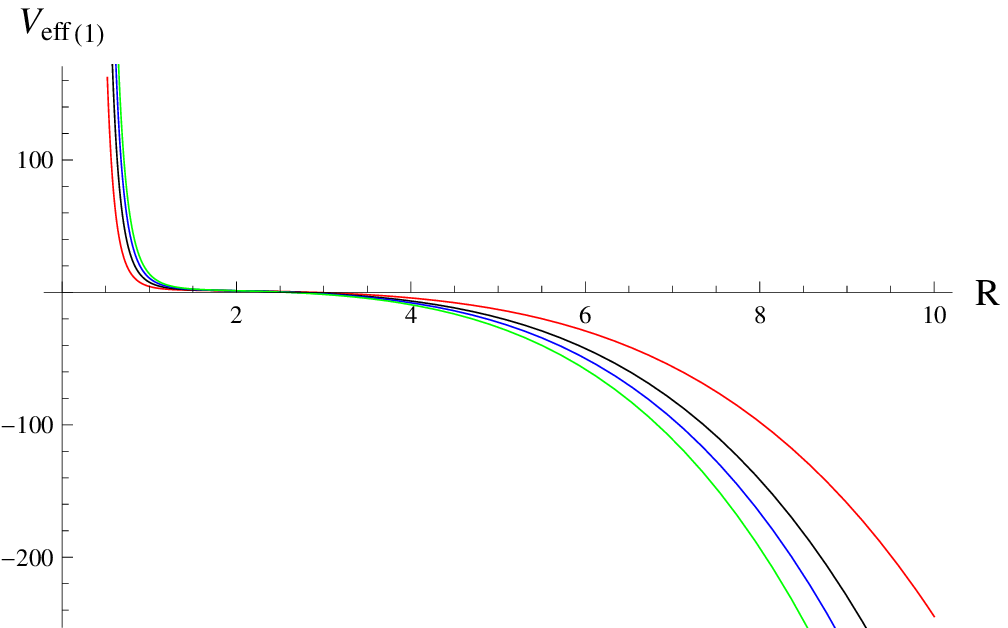,width=.42\linewidth}\epsfig{file=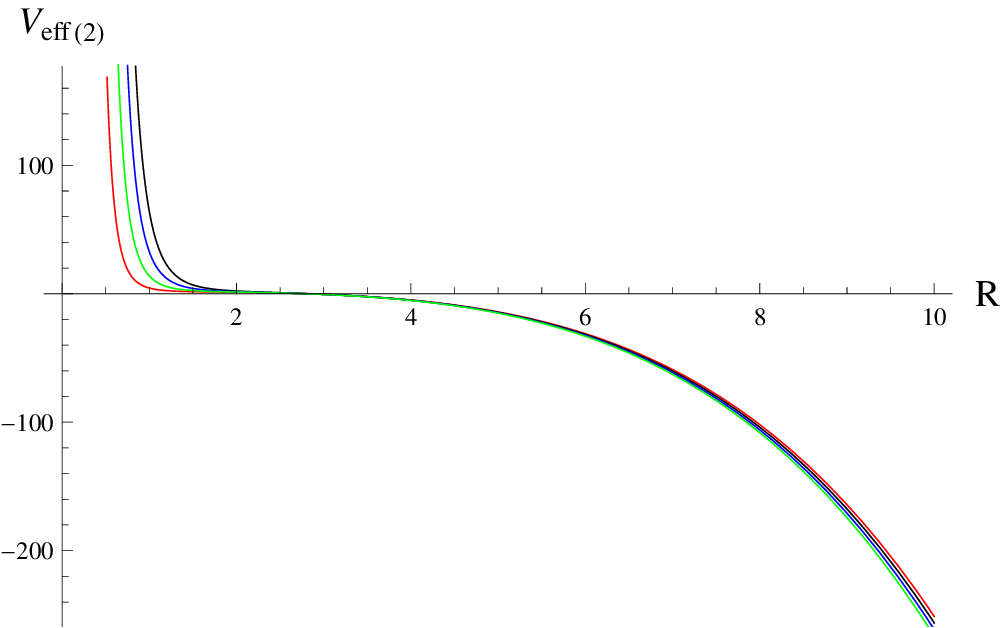,width=.42\linewidth}
\\\epsfig{file=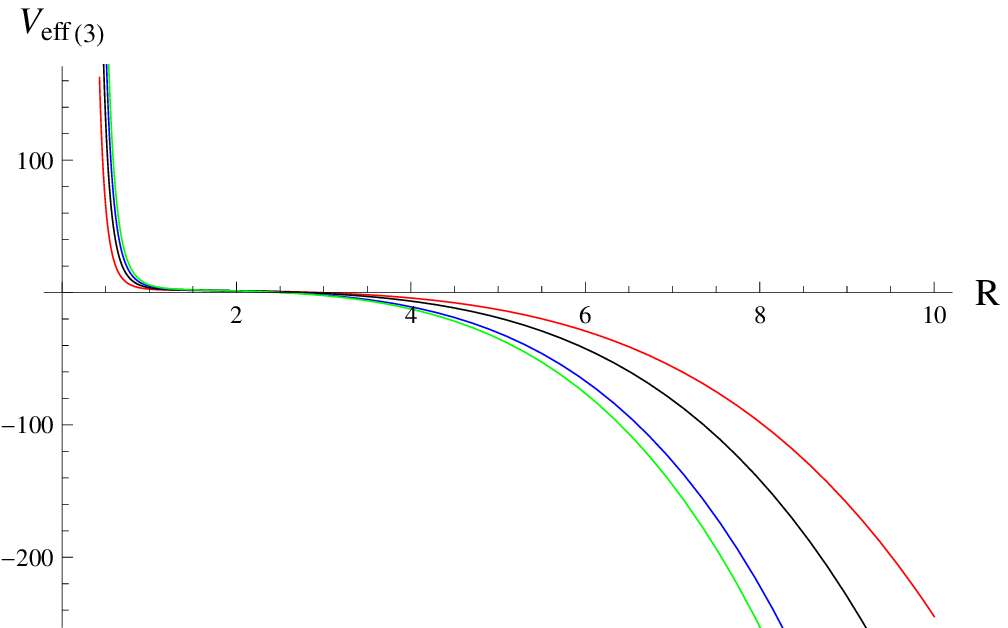,width=.42\linewidth}
\caption{Behavior of $V_{eff}$ of the massless shell by varying
$\lambda$.}
\end{figure}

In Figures \textbf{6}-\textbf{11}, we examine numerical results for
the massless scalar field using $M_{-}=0$, $M_{+}=1$, $R_{0}$, $e
=1$ and $\lambda=1$ . Figure \textbf{6} shows that increasing and
decreasing shell radius lead to expansion and collapse,
respectively. The behavior of the effective potential is presented
in Figure \textbf{7} which is divided into two regions. The upper
region has a positive potential that leads to expansion. There is a
turning (saddle) point where $V_{eff}=0$, the shell stops for a
while and then changes its behavior at $R=3.5$ for Bardeen and
Hayward BHs while at $R\approx4$ for ABGB BH. The effective
potential decreases infinitely after these points and becomes
negative. This suggests that for large values of $R$ the shell
begins to contract continuously. Figures \textbf{8} and \textbf{9}
describe behavior of the effective potential by varying $M_{+}$ and
$M_{-}$. The saddle point ($V_{eff}=0$) separates shell's motion
into two regions: the upper (positive) and lower (negative) regions
describe expansion and contraction of the shell, respectively.
Figures \textbf{10} and \textbf{11} represent the behavior of
effective potential by varying charge and $\lambda$. Again, the
shell depicts three types of motion, expands in the upper region
($V_{eff}>0$), in equilibrium position ($V_{eff}=0$) and in the
lower region the effective potential diverges negatively which leads
the shell to collapse.

\subsubsection{Massive Scalar Shell}

Here we discuss the case when the thin-shell is composed of a
massive scalar field with scalar potential. From Eq.(\ref{20}), we
obtain
\begin{equation}\label{26}
\dot{\varphi}^{2}=\rho+p,\quad V(\varphi)=\frac{1}{2}(p-\rho).
\end{equation}
We take $p$ as an explicit function of $R$, i.e., $p=p_{0}
e^{-\check{k} R}$, where $\check{k}$ and $p_{0}$ are constants.
Using the value of $p$ in Eq.(\ref{11}), we find
\begin{equation}\label{27}
\rho=\frac{\zeta}{R^{2}}+\frac{2(1+\check{k}R)p_{0}e^{-\check{k}
R}}{\check{k}^{2}R^{2}},
\end{equation}
where $\zeta$ is the constant of integration. Inserting the values
of $p$ and $\rho$ in Eq.(\ref{26}), we obtain
\begin{eqnarray}\label{28}
V(\varphi)&=&\frac{\zeta}{2R^{2}}-\frac{p_{0}e^{-\check{k}
R}}{2}\left(1-\frac{2(1+\check{k}R)}{\check{k}^{2}R^{2}}\right),
\\\label{29} \dot{\varphi}^{2}&=&\frac{\zeta}{R^{2}}+p_{0}e^{-\check{k}
R}\left(1+\frac{2(1+\check{k}R)}{\check{k}^{2}R^{2}}\right),
\end{eqnarray}
which satisfy the KG equation. Using Eqs.(\ref{27})-(\ref{29}) in
(\ref{23}), it follows that
\begin{eqnarray}\nonumber
V_{eff_{(1)}}(R)&=&1-\left(\frac{2R^{3}}{(R^{2}+
e^{2})^{\frac{3}{2}}}\right)^{2}\left(\frac{M_{+}-M_{-}}{m}\right)^{2}
-\frac{(M_{+}+M_{-})R^{2}}{(R^{2}+e^{2})^{\frac{3}{2}}}\\\label{30}&-&\left(\frac{m}{2R}\right)^{2},
\\\nonumber
V_{eff_{(2)}}(R)&=&1-\left(\frac{2R^{3}}{R^{3}+
2e^{2}}\right)^{2}\left(\frac{M_{+}-M_{-}}{m}\right)^{2}
-\frac{(M_{+}+M_{-})R^{2}}{(R^{3}+
2e^{2})}\\\label{t}&-&\left(\frac{m}{2R}\right)^{2},
\\\nonumber
V_{eff_{(3)}}(R)&=&1-\left(\frac{2R^{3}}{(R^{2}+
e^{2})^{\frac{3}{2}}}\right)^{2}\left(\frac{M_{+}-M_{-}}{m}\right)^{2}
-\frac{(M_{+}+M_{-})R^{2}}{(R^{2}+e^{2})^{\frac{3}{2}}}\\\label{ttt}&+&
\frac{e^{2}R^{2}}{(R^{2}+e^{2})^{2}}-\left(\frac{m}{2R}\right)^{2},
\end{eqnarray}
where
\begin{equation}
m=4\pi R^{2}\rho\equiv4\pi\zeta+\frac{8\pi
p_{0}e^{-\check{k}R}}{\check{k}^{2}}(1+\check{k}R).
\end{equation}
\begin{figure}\centering
\epsfig{file=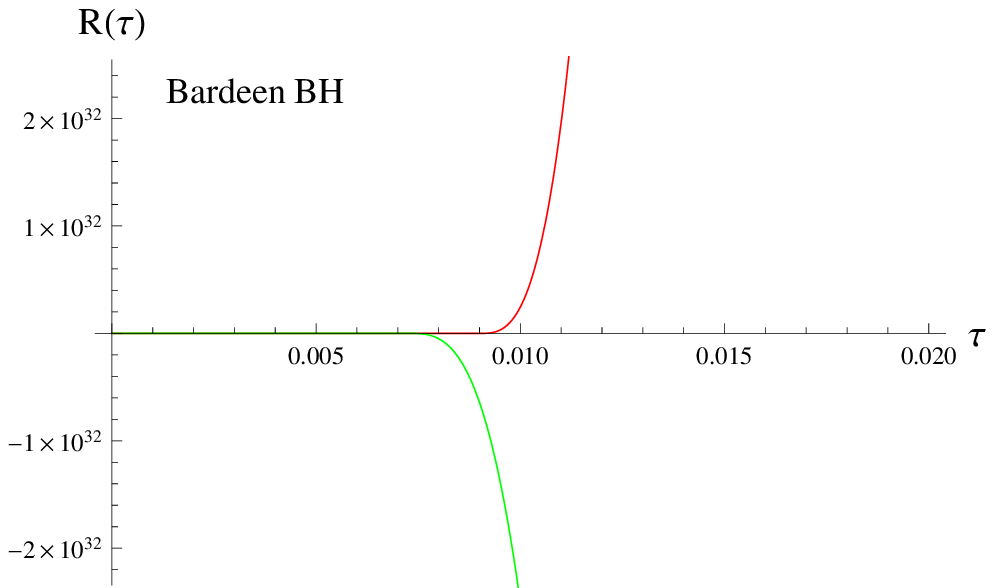,width=.42\linewidth}\epsfig{file=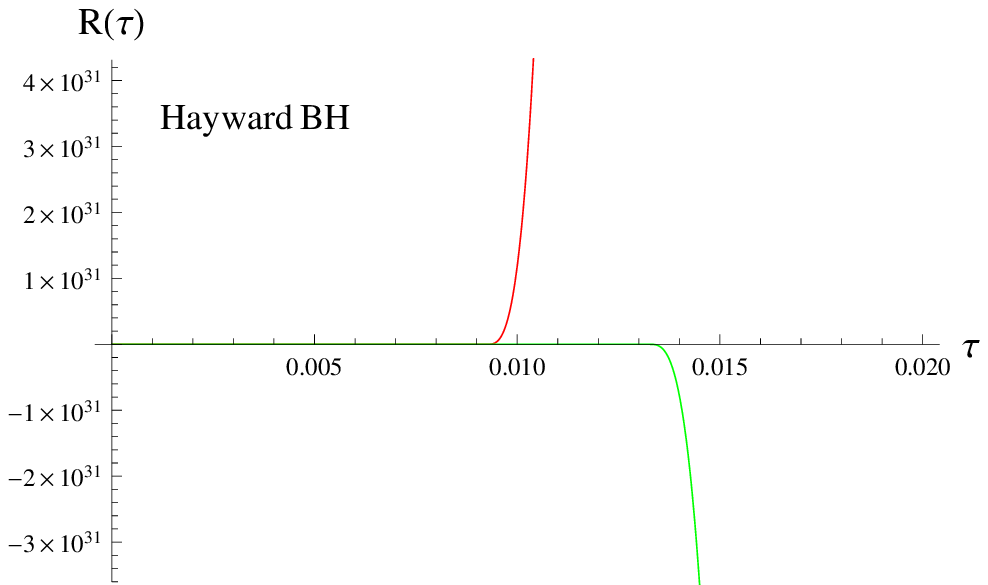,width=.42\linewidth}
\\\epsfig{file=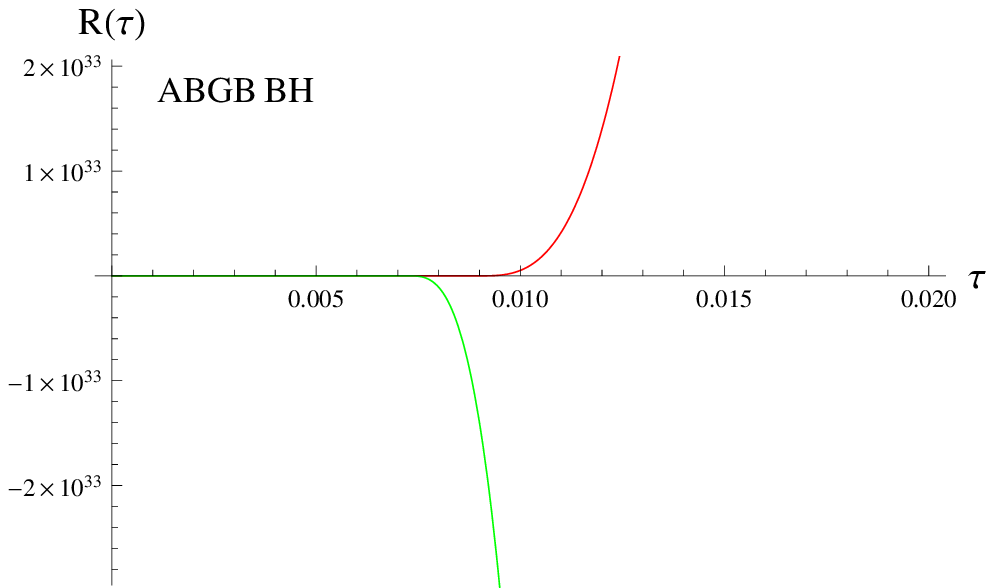,width=.42\linewidth}
\caption{Plots of the shell radius in massive scalar field.}
\end{figure}
\begin{figure}\centering
\epsfig{file=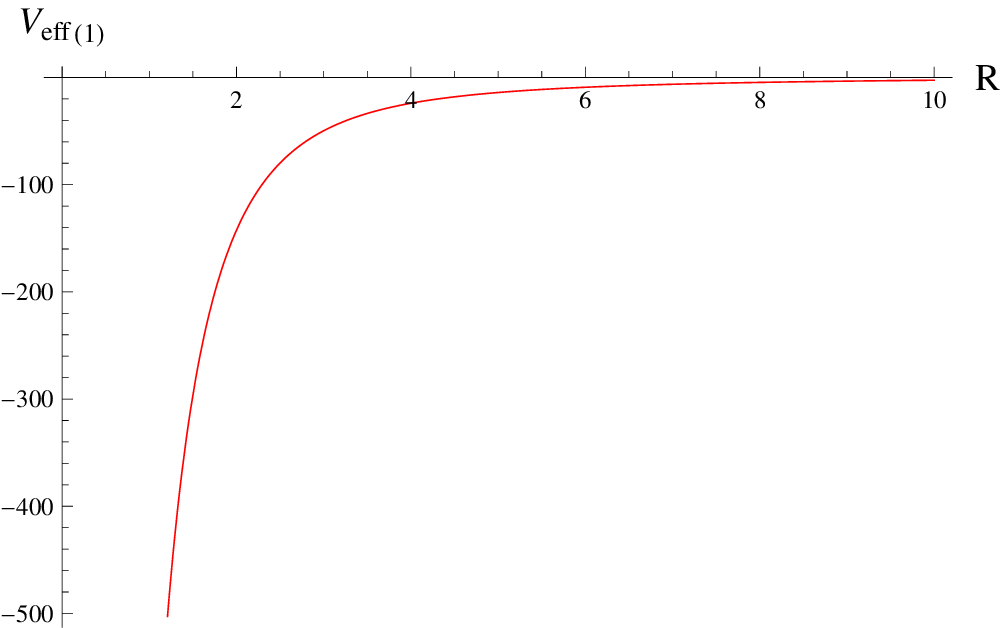,width=.42\linewidth}\epsfig{file=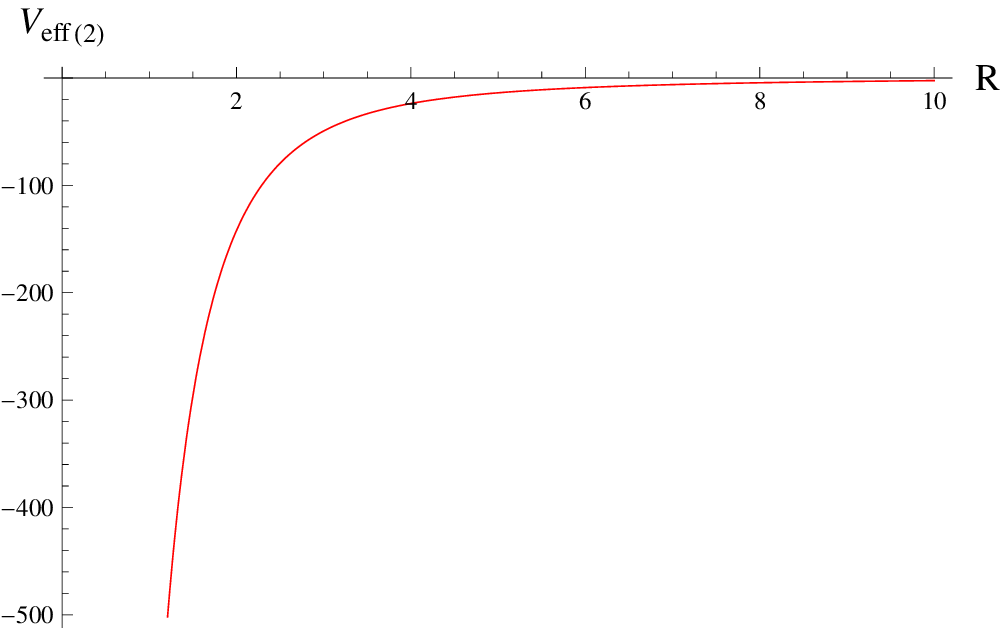,width=.42\linewidth}
\\\epsfig{file=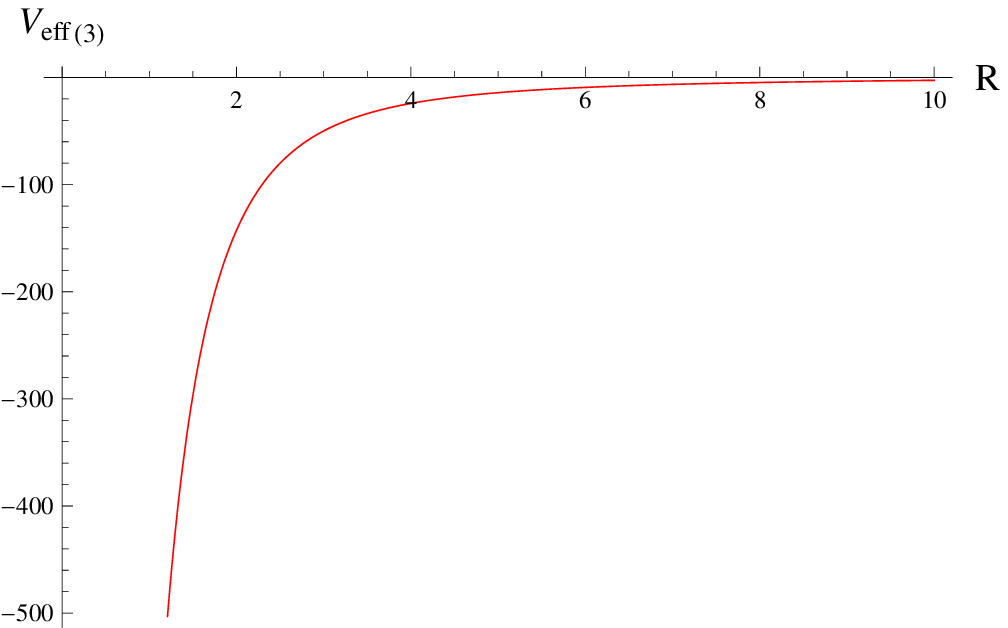,width=.42\linewidth}
\caption{Behavior of $V_{eff}$ of a massive scalar shell. The plots
for Bardeen (left) and Hayward (right) is given in the upper panel
while the graph of ABGB BH is in the lower panel.}
\end{figure}
\begin{figure}\centering
\epsfig{file=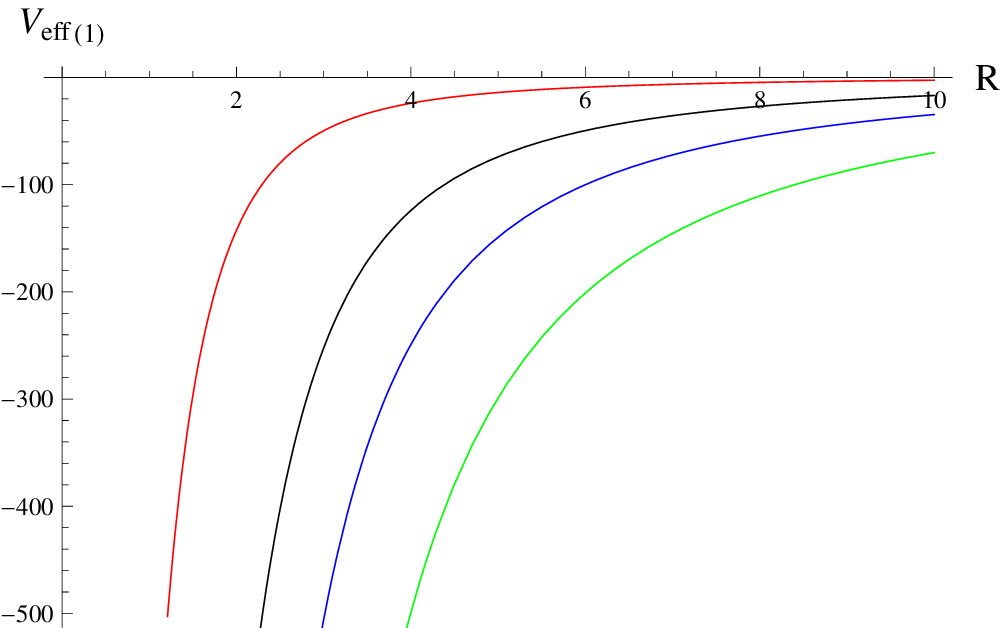,width=.42\linewidth}\epsfig{file=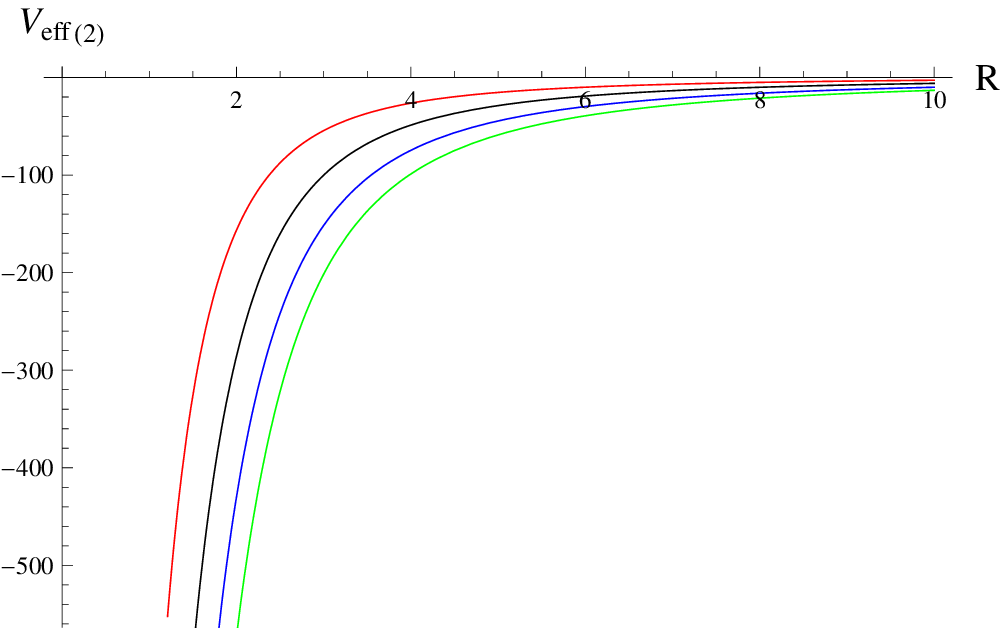,width=.42\linewidth}
\\\epsfig{file=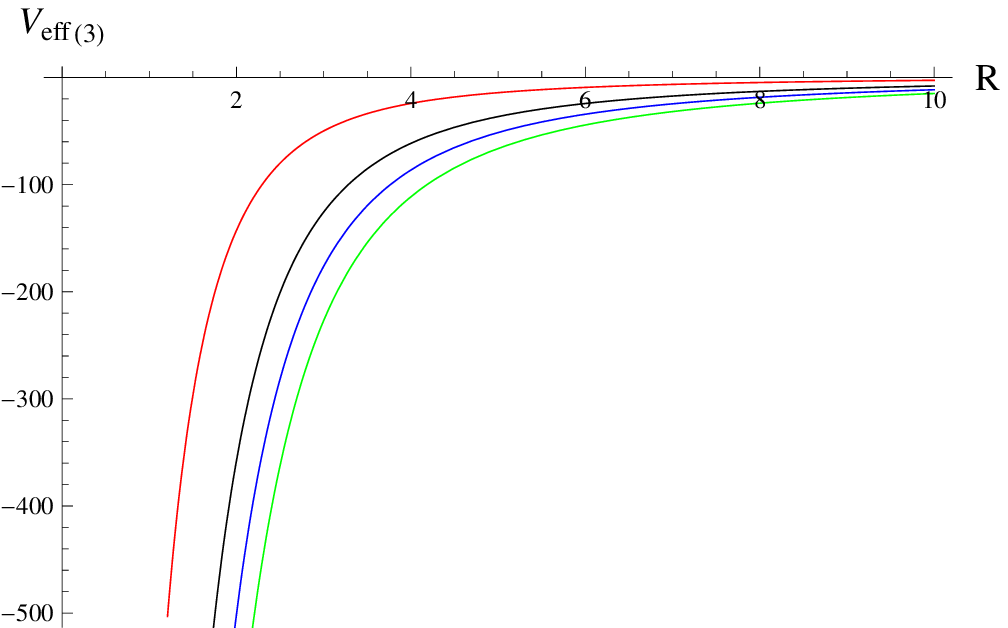,width=.42\linewidth}
\caption{plots of $V_{eff}$ of the massive scalar shell by using
different values of $e$. }
\end{figure}

Figures \textbf{12}-\textbf{14} show the behavior of thin-shell for
the massive scalar field for $M_{-}=0$, $M_{+}=1$,
$R_{0}=p_{0}=\check{k}=1$, $e = 1$ and $\zeta=3$. Figure \textbf{12}
describes the nature of the shell radius for the massive scalar
field. The upper curve corresponds to constant motion which leads to
expansion of the shell with the increasing time. The lower curve
follows the same initial configuration leading the shell to
collapse. Figures \textbf{ 13} and \textbf{ 14} illustrate the
behavior of effective potential for the massive scalar field with
fixed mass and varying the charge parameter. In Figure \textbf{13},
we see that the effective potential diverges for the initial data.
This negative effective potential indicates that the gravitational
forces lead the shell to collapse. In Figure \textbf{14}, the
effective potential is plotted for different values of charge
parameter which shows that the shell collapses for all values of
charge.

\section{Final Remarks}

In this paper, we have examined the dynamics of spherically
symmetric scalar thin-shell (both massless and massive scalar
fields) by taking regular geometry for the interior as well as
exterior regions. We have considered three regular BHs, Bardeen,
Hayward and ABGB. The equation of motion (\ref{15}) and the KG
equation (\ref{22}) can completely describe the dynamical behavior
of the shell. We have discussed solutions of these equations
graphically shown in Figures \textbf{1}-\textbf{5} which represent
both collapse and expansion of the shell. The charged and uncharged
shell provide a comparison between the shell's motion of singular
and non-singular BHs, respectively. Initially the velocity of the
shell with respect to stationary observer in regular spacetime is
slower as compared to the Schwarzschild case (Figures:
\textbf{1}-\textbf{3}). It is found that the scalar field increases
for the case of collapse while for expansion it shows decreasing
behavior for all BHs (Figures: \textbf{4}-\textbf{5}).

In massless case, the increase or decrease in shell's radius along
the proper time represents that the shell expands continuously or
collapses (Figure \textbf{6}). The motion of the shell is determined
by the effective potential (Figures \textbf{7}-\textbf{11}). The
shell is partitioned into two regions by a saddle point
($V_{eff}=0$) which expands forever in the region $V_{eff}>0$ while
the region with $V_{eff}<0$ indicates the collapsing shell. For
massive scalar field, the behavior of the shell radius shows that it
either expands forever or undergoes collapse (Figure \textbf{12}).
The effective potential (Figure \textbf{13} and \textbf{14}) is
always negative for the fixed mass (interior and exterior) with
different values of $e$. These results indicate that the massive
scalar shell always collapse to zero size for the considered
parameters. We conclude that there are three possibilities in the
dynamical evolution of the scalar thin-shell: continuous expansion
($V_{eff}>0$), stable configuration ($V_{eff}=0$) and gravitational
collapse ($V_{eff}<0$).

\end{document}